\documentclass[amsmath, amsfonts, twocolumn, superscriptaddress, prx]{revtex4-1}
\usepackage{graphicx}
\usepackage{epsfig}
\usepackage{bm}
\usepackage{dcolumn}
\usepackage{amsmath}
\usepackage{amssymb}
\usepackage{color}
\usepackage{pifont}
\usepackage[varg]{txfonts}

\newcommand{\Det}{\mathop{\rm Det}\nolimits}
\newcommand{\Tr}{\mathop{\rm Tr}\nolimits}
\newcommand{\tr}{\mathop{\rm tr}\nolimits}
\newcommand{\dg}{^\dagger}
\newcommand{\up}{\uparrow}
\newcommand{\dn}{\downarrow}

\begin{document}

\title{Anomalous thermodynamic properties of quantum critical superconductors}

\author{Maxim Khodas}
\affiliation{Racah Institute of Physics, Hebrew University of Jerusalem, Jerusalem 91904, Israel}

\author{Maxim Dzero}
\affiliation{Department of Physics, Kent State University, Kent, Ohio 44242, USA}

\author{Alex Levchenko}
\affiliation{Department of Physics, University of Wisconsin--Madison, Madison, Wisconsin 53706, USA}

\begin{abstract}
Recent high-precision measurements employing different experimental techniques have unveiled an anomalous peak in the doping dependence of the London penetration depth which is accompanied by anomalies in the heat capacity in iron-pnictide superconductors at the optimal composition associated with the hidden antiferromagnetic quantum critical point. We argue that finite temperature effects can be a cause of observed features. Specifically we show that quantum critical magnetic fluctuations under superconducting dome can give rise to a nodal-like temperature dependence of both specific heat and magnetic penetration depth in a fully gapped superconductor. In the presence of line nodes in the superconducting gap fluctuations can lead to the significant renormalization of the relative slope of $T$-linear penetration depth which is steepest at the quantum critical point. The results we obtain are general and can be applied beyond the model we use. 
\end{abstract}

\date{October 31, 2020}

\maketitle

\section{Introduction} 

\begin{figure*}
  \centering
 \includegraphics[width=7in]{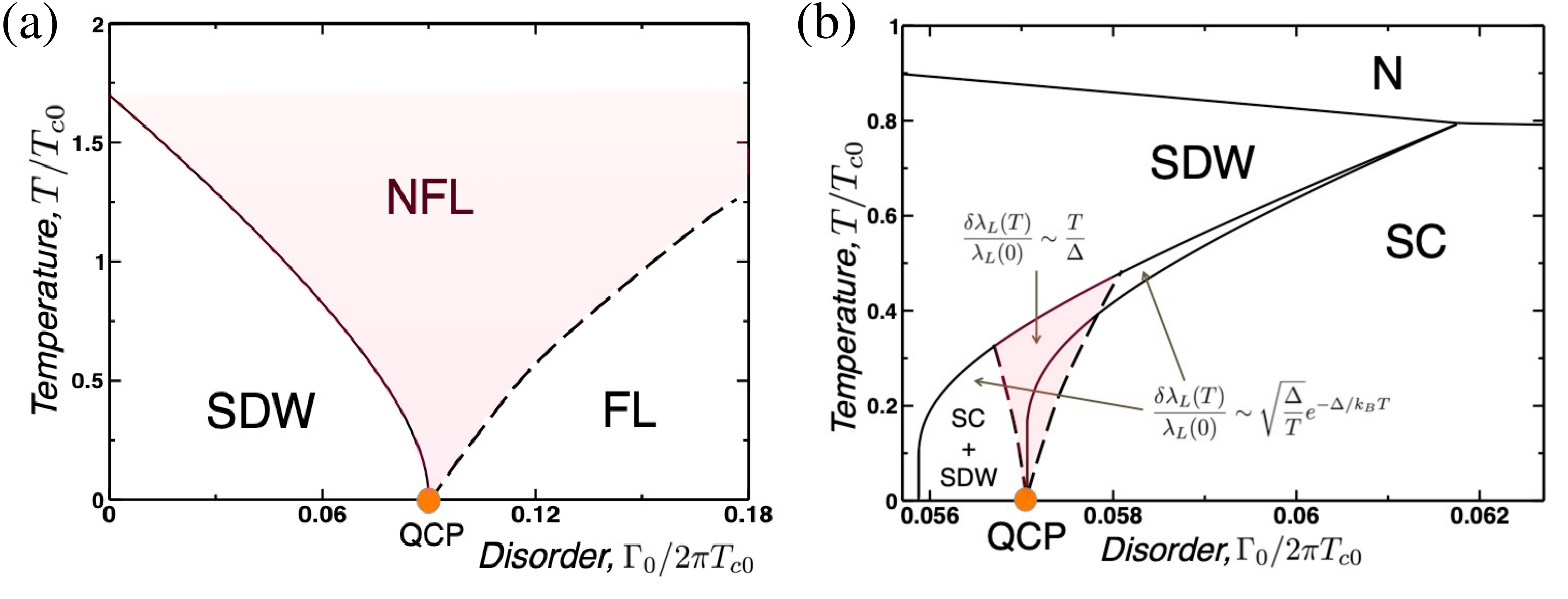}
  \caption{Phase diagram of the (a) magnetic SDW quantum criticality without SC and (b) with SC coexistence computed numerically from the so-called disorder model presented in Sec \ref{Sec:Model}. In the superconducting case the fan region extending away from the QCP represents an anomalous part of the phase diagram where nodal-like behavior of gapped fermions emerges.}\label{Fig-QCP-PD}
\end{figure*}

Quantum phase transitions and quantum criticality are among the central concepts in the physics of correlated electrons \cite{Sachdev-Book,Vojta-Review}. In general, quantum fluctuations (QF) near magnetic e.g. spin-density-wave (SDW) quantum critical point (QCP) give rise to non-Fermi liquid behavior that manifests in singularities and nonanalyticity of various electronic characteristics \cite{Stewart,Abanov,Metlitski,Sur-Lee}. This problem is further enriched in the situations when magnetic instability competes with superconductivity (SC) \cite{Pelissetto,Basov,Chubukov-Review,Efetov,Chubukov-Eliashberg}, see Fig. \ref{Fig-QCP-PD} for the exemplary phase diagram. This is the case in the context of cuprate- and iron-based superconductors where interest in the topic is constantly fueled by a multitude of experimental activities (for the recent detailed reviews see, e.g., Refs. \cite{Shibauchi-Review,Taillefer-Review} and references herein).

The key signatures of QCP behavior in SCs include correlated anomalies in both transport coefficients and thermodynamic properties, which emerge in different temperature regimes of the phase diagram when the system is tuned by an external control parameter (e.g., doping $x$) to an optimal composition $x_c$. Indeed, some of these anomalies persist in the normal state such as linear-in-$T$ resistivity observed in various materials at $x_c$ \cite{Mackenzie-Planckian,Shekhter,Analytis-Planckian,Taillefer-Planckian,Paglione-Planckian}. It is typically accompanied by anomalies in the transverse Hall and thermoelectric responses \cite{Analytis-Hall,Taillefer-ThHall,Canfield-TEP}. When system is brought to the proximity of the phase transition, then thermally activated fluctuations of magnetic and superconducting orders start to play a dominant role. This translates into the nonmonotonic discontinuity of the specific heat jump which also peaks at $x_c$ \cite{Hardy,Carrington,Klauss}. Further, when the system is cooled into the superconducting state, quantum fluctuations proliferate and their effect becomes most pronounced near the transition line that separates pure superconducting and mixed phase coexisting with magnetism that ultimately terminates at the QCP. Near that region one often detects enhanced  critical supercurrents \cite{Eisterer} and observes the apparent sharp peak in the magnetic penetration depth  \cite{Matsuda,Auslaender-1,Auslaender-2,Zheng,Prozorov}. 

In part motivated by these results, the interplay of possible magnetic and structural quantum phase transitions shielded by the superconductivity was a subject of an immediate scrutiny \cite{Fernandes-QCP}. In a parallel vein of studies, various models of Planckian resistivity were proposed \cite{Efetov-T-linear,Balents,Senthil,Patel}, thermal and electrical transport properties across antiferromagnetic quantum transition were considered \cite{Eberlein}, and further extensions to anomalous Hall phenomena were developed \cite{Vafek,SLAL}. Thermodynamic signatures of QCP were analyzed theoretically in the context of the specific heat \cite{Vavilov,Kuzmanovski,Carvalho} and Josephson effect \cite{Dzero-Josephson,Kirmani}. However, despite much of the efforts \cite{LVKC,Chowdhury-PRL,Nomoto-Ikeda,Chowdhury-PRB,Dzero,Zhang} there is no consensus on the explanation of the observed peak in the London penetration depth. 

In this work we show that finite-temperature effects of quantum spin-density-wave fluctuations yield anomalous thermodynamic properties of gapped fermions with pronounced power-law dependencies in both specific heat and London penetration depth which is reminiscent of that of nodal superconductors. We also demonstrate generality of these results, in particular robustness to effects of disorder. 

The rest of the paper is organized as follows. In Sec. \ref{Sec:Model} we present a disorder model of SC-SDW coexistence, develop its mean-field description, and establish a phase diagram. In Sec. \ref{Sec:Applications} we apply this model to calculate quantum fluctuation corrections to heat capacity and penetration depth near magnetic QCP hidden beneath the SC dome. We summarize our main results in Sec. \ref{Sec:Summary}. The main sections of the paper are accompanied by several Appendices \ref{App:QuasiClass}--\ref{App:Fluctuations}, where an in-depth technical discussion is provided and supporting detailed calculations are carried out. Throughout the paper we work in the units $\hbar=k_B=c=1$.  

\section{Disorder model of SC-SDW coexistence}\label{Sec:Model}

\subsection{Mean-field description}

We adopt the two-band model which is defined by the Hamiltonian \cite{Vavilov,Dzero}:
\begin{equation}\label{Eq:H}
\hat{H}=\hat{H}_0+\hat{H}_{\textrm{sdw}}+\hat{H}_{\textrm{sc}}+\hat{H}_{\textrm{dis}}.
\end{equation} 
The first term describes noninteracting fermions occupying two (one electron- and one hole-like) bands: 
\begin{equation}\label{Eq:H0}
\hat{H}_0=\sum_{\bm{k}}\xi_{\bm{k}}\Psi_{\bm{k}}^\dagger (\hat{\tau}_3 \hat{\rho}_3\hat{\sigma}_0)\Psi_{\bm{k}},
\end{equation} 
where we take simple parabolic band dispersion $\xi_{\bm{k}}=k^2/2m-\mu$, defined relative to the chemical potential $\mu$, and all momenta are counted relative to the center of the corresponding pocket. In the Balian-Werthammer representation \cite{BW}, the eight component spinor 
\begin{equation}\label{Eq:Psi}
\Psi^\dag_{\bm{k}}=(\hat{c}_{\bm{k}\uparrow}^\dag, ~\hat{c}_{\bm{k}\downarrow}^\dag,~\textrm{-}\hat{c}_{-\bm{k}\downarrow}, ~\hat{c}_{-\bm{k}\uparrow},
~\hat{f}_{\bm{k}\uparrow}^\dag,~\hat{f}_{\bm{k}\downarrow}^\dag,~\textrm{-}\hat{f}_{-\bm{k}\downarrow}, ~\hat{f}_{-\bm{k}\uparrow})
\end{equation}
is composed of electron-$c$ and hole-$f$ creation and annihilation operators with spin projections $\uparrow\downarrow$. Three sets of Pauli matrices $(\hat{\tau},\hat{\rho},\hat{\sigma})$ operate in the band, Nambu, and spin spaces, respectively. The second term in Eq. \eqref{Eq:H} describes magnetic interpocket interaction between fermions 
\begin{equation}\label{Eq:Hsdw}
\hat{H}_{\textrm{sdw}}=-\frac{g_{\textrm{sdw}}}{2}\sum_{\bm{Q}}\bm{S}_{\bm{Q}}\bm{S}_{-\bm{Q}},
\end{equation} 
where the magnetization fluctuation at momentum $\bm{Q}$ is $\bm{S}_{\bm{Q}}=(1/2)\sum_{\bm{k}}\Psi^\dag_{\bm{k}+\bm{Q}}\hat{\bm{\Xi}}\Psi_{\bm{k}}$,  $\hat{\bm{\Xi}}= \hat{\tau}_1\hat{\rho}_0\hat{\bm{\sigma}}$.
The third term in Eq. \eqref{Eq:H} captures pairing interaction and in the model of $s^{\pm}$ order parameter changing sign between the hole and electron pockets takes the form
\begin{equation}\label{Eq:Hsc}
\hat{H}_{\textrm{sc}}=-\frac{g_{\textrm{sc}}}{2}\sum_{\bm{k}\bm{k}'}B_{\bm{k}}B_{\bm{k}'}, 
\end{equation}
where the fermion bi-linear is defined as $B_{\bm{k}}=\Psi^\dag_{\bm{k}}(\hat{\tau}_3\hat{\rho}_1\hat{\sigma}_0)\Psi_{\bm{k}}$. 
In this low-energy description we impose high-energy cutoff $\Lambda$, and consider only angle-independent interactions in the SDW channel and in the $s^{\pm}$ SC channel with the couplings $g_{\textrm{sdw}}$ and $g_{\textrm{sc}}$. With the last term in Eq. \eqref{Eq:H} we introduced a disorder potential into the problem. We account for two types of scattering processes: the intraband disorder with potential $U_0$, which scatters quasiparticles within the same band, and interband scattering between the Fermi pockets mediated by the potential $U_\pi$.
In the basis of spinors $\Psi_{\bm{k}}$ the disorder term reads 
\begin{equation}\label{Eq:Hdis}
\hat{H}_{\text{dis}}=\!\!\sum_{\bm{k}\bm{k}'\bm{R}_j}\!\! \Psi^\dag_{\bm{k}}\left[U_0(\hat{\tau}_0\hat{\rho}_3\hat{\sigma}_0)+U_\pi(\hat{\tau}_1 \hat{\rho}_3\hat{\sigma}_0)\right]\Psi_{\bm{k}'}e^{i(\bm{k}-\bm{k}')\bm{R}_j}
\end{equation}
 where summation goes over the random locations $\bm{R}_j$ of individual impurities. When performing disorder averaging within the self-consistent Born approximation we assume that concentration of impurities is $n_{\text{imp}}$. This naturally introduces two scattering rates into the model $\Gamma_{0,\pi}={\pi\nu_F n_{\text{imp}}}|U_{0,\pi}|^2/4$, where $\nu_F$ is the total quasiparticle density of states at the Fermi energy. 

The mean-field (MF) analysis of this model proceeds in a standard way by decoupling interaction terms via Hubbard-Stratonovich transformation with magnetic $M$ and superconducting $\Delta$ order parameters, and integrating out fermions \cite{Vavilov,Dzero}. This approach naturally leads to the semiclassical description based on the Eilenberger equation, which is further elaborated on in Appendix \ref{App:QuasiClass}. 

In this treatment, the pure SC transition temperature $T_c$ is suppressed only by the interband scattering as described in accordance with the Abrikosov-Gor'kov scenario 
\begin{equation}
\ln\left(\frac{T_{c0}}{T_c}\right)=\psi\left(\frac{1}{2}+\frac{\Gamma_\pi}{\pi T_c}\right)-\psi\left(\frac{1}{2}\right),
\end{equation} 
where $T_{c0}\simeq \Lambda e^{-2/\nu_Fg_{\text{sc}}}$ and $\psi(x)$ is the digamma function. This is similar to the equation for $T_c$ in conventional single-band $s$-wave superconductors with magnetic impurities, and in the unconventional $d$-wave superconductors with potential impurities. In contrast, pure SDW transition temperature $T_s$ is suppressed by the total scattering rate, 
\begin{equation}
\ln\left(\frac{T_{s0}}{T_s}\right)=\psi\left(\frac{1}{2}+\frac{\Gamma_0+\Gamma_\pi}{\pi T_s}\right)-\psi\left(\frac{1}{2}\right),
\end{equation} 
where $T_{s0}\simeq \Lambda e^{-2/\nu_Fg_{\text{sdw}}}$. As a result of different sensitivity to disorder, there exists a finite parameter range in $\Gamma_{0,\pi}$ where both orders $M$ and $\Delta$ can coexist simultaneously. The magnetic QCP is defined by the condition $T_s(\Delta)=0$, which corresponds to $M=0$ for certain values of $\Gamma_{0,\pi}$, see Fig. \ref{Fig-QCP-PD}(b) for an example. We note that this phase diagram was calculated numerically for the choice of parameters when $\Gamma_\pi/\Gamma_0=0.325$ and $T_{s0}/T_{c0}=1.7$.

At this point, it is worth commenting that usually stability of QCP in disordered systems is analyzed through the prism of the Harris criterion \cite{Harris,Chayes}, namely when disorder is introduced on top of the control parameter that defines QCP. In the model considered here, it is disorder itself that defines QCP and, as we show below, controls fluctuations around it. 

\subsection{SDW fluctuation propagator in SC state} 

Extending theory beyond the mean field description we consider the critical fluctuations that mediate an effective interaction in the spin channel $(S_z)$ represented by the propagator
\begin{equation}\label{Eq:L}
L_{Q,\Omega_m}=\left( g_{\text{sdw}}^{-1}+\Pi^{z}_{Q,\Omega_m}\right)^{-1}.
\end{equation} 
The disorder-averaged polarization operator $\Pi^z_{Q,\Omega_m}$ needs to be calculated by resumming the whole sequence of ladder-type diagrams with impurity line insertions into the fermionic loop. It can be shown that, in the proper matrix basis representation, this averaging can be reduced to a geometric series that sums to    
\begin{equation}\label{Eq:Pi} 
\Pi^z_{Q,\Omega_m}=T\sum_{\omega_n}\left[\hat{P}_{Q,\Omega_m}\left(1-\hat{\Gamma}\circ\hat{P}_{Q,\Omega_m}\right)^{-1}\right]_z,
\end{equation}
where the notation with subscript $z$ implies a specific matrix element. The convolution in Eq. \eqref{Eq:Pi} between the disorder matrix ($\hat{\Gamma}$) and matrix polarization function ($\hat{P}_{Q,\Omega_m}$) requires a specification of the basis representation which is explained in Appendix \ref{App:Disorder} [see Eqs. \eqref{dec17a} and \eqref{Eq:Pz-vs-P103} for definitions]. For instance, the diagonal matrix element of the bare fermionic loop, namely the polarization operator without vertical impurity lines, is given by   
\begin{equation}\label{Eq:Pi-bare} 
P^z_{Q,\Omega_m} (\omega_n)= \sum_k\tr\left[\hat{\Xi}^z \hat{G}_{k_+,\omega_+} \hat{\Xi}^z \hat{G}_{k_-,\omega_-}\right],
\end{equation}
with $\hat{\Xi}^z=\hat{\tau}_1\hat{\rho}_0\hat{\sigma}_3$, $k_{\pm}=k\pm Q/2$, and $\omega_\pm=\omega_n\pm\Omega_m/2$. 
The latter is defined via the disorder averaged single-particle propagator Green's function
\begin{equation}
[\hat{G}_{k,\omega_n}]_{\alpha \beta} = -\int_0^{T^{-1}} d \tau e^{ i \omega_n \tau} \langle \Psi_{k\alpha}(\tau) \Psi^{\dag}_{k\beta}\rangle.
\end{equation}
The trace in Eq. \eqref{Eq:Pi-bare} assumes reduction over all the matrix indices. Later in the text we will use a global trace that in addition includes summation over the Matsubara frequency, $\omega_n=\pi T(2n+1)$, and momenta, thus introducing a notation $\Tr[\ldots]=T\sum_{k,\omega}\tr[\ldots]$. 

The critical paramagnon described by the spin correlation function in Eq. \eqref{Eq:L} with $\Omega_m=2\pi mT$ softens towards the QCP, 
\begin{equation}
g_{\text{sdw}}^{-1}+\Pi^z_{Q,\Omega_m} =\pi \nu_F \left(\gamma +   Q^2/Q_c^2 + \Omega_m^2/\Omega_c^2 \right),
\end{equation}
reached at $\Gamma_0 = \Gamma_c $ such that,
$\gamma(\Gamma) \approx \gamma'_{\pm}|\Gamma - \Gamma_c|$, $\gamma'_{\pm} = |d \gamma/d\Gamma|$
taken at $\Gamma = \Gamma_c \pm 0^+$. It should be noted that in a SC state, the dynamical exponent due to the exchange of near-critical SDW fluctuations changes from $z=2$ to $z=1$ as compared to the normal case, because fermions which contribute to bosonic dynamics become massive excitations protected by a gap. This is reflected in the asymptotic expansion of $\Pi^z_{Q,\Omega_m}$ having a $\Omega^2_m$ term, which is valid at low energies $\{v_FQ,\Omega_m\}\ll\Delta$.  

We find in this model rather generally that the QCP is located at $\Gamma_c=2\pi aT_{c0}$ where a precise numerical value of the parameter $a(\Gamma_\pi/\Gamma_0,T_{c0}/T_{s0})$ depends on the choice of two ratios between scattering rates and bare interaction parameters (or alternatively bare transition temperatures). Furthermore, while the ratio $\gamma'_{+}/\gamma'_-$ can be arbitrary, the low-energy expansion coefficients $Q_c$ and $\Omega_c$ may be computed right at the QCP. Further details and generalities of calculation of $L_{Q,\Omega_m}$ are relegated to Appendices \ref{App:Disorder} and \ref{App:Propagator}, where in particular we discuss separately disorder renormalizations of vertex functions and impurity ladders in the low-energy expansion of the polarization operator defined by Eq. \eqref{Eq:Pi}.

\section{Applications}\label{Sec:Applications}

\subsection{Specific heat near QCP} 

We now focus on the impact of quantum SDW fluctuations on the low-temperature behavior of the specific heat inside the dome of $s^{\pm}$ superconductivity. Recall that at the level of the mean-field analysis, the low-$T$ asymptotic behavior of the specific heat in a fully gapped SC state is exponential $C_{\text{MF}}\propto (\Delta/T)^{3/2}e^{-\Delta/T}$ for $T\ll\Delta$. Our intent is to investigate the fate of this result as the system approaches a QCP by accounting for the extra contribution of the spin fluctuations. Following the standard procedure, we integrate out these soft magnetic modes from the action. At the Gaussian level we thus get a renormalized free energy of a superconductor per unit layer area $F=F_{\text{SC}}(\Delta,M)+\delta F_{\text{QF}}$ that can be expressed in terms of the SDW propagator from Eq. \eqref{Eq:L} as follows 
\begin{equation}\label{Eq:FQF}
\frac{\delta F_{\text{QF}}}{T}=\frac{\mathcal{N}}{2}\Tr\ln\left[L^{-1}_{Q,\Omega_m}\right],
\end{equation}
where $\mathcal{N}$ counts the number of soft modes.
In our model $\mathcal{N}=3$ at $x>x_c$ and $\mathcal{N}=1$ at $x<x_c$ as only the longitudinal mode has a mass changing with $\Gamma$. The factor $1/2$ is present because the paramagnons commensurate with the lattice are represented by a real boson field. Next, performing the Matsubara sum, and separating the temperature independent term, one can easily analyze limiting cases for the corresponding specific heat correction $\delta C_{\text{QF}}=-T\partial^2_T \delta F_{\text{QF}}$ (see Appendix \ref{App:Fluctuations} for further technical details). We thus find in a broad regime of temperatures  $\Delta_{\text{QCP}}<T<\Delta$, 
\begin{equation}\label{C-QF}
\delta C_{\text{QF}}=\frac{9\zeta(3)}{\pi}\left(\frac{v_FQ_c}{\Omega_c}\right)^2\left(\frac{T}{v_F}\right)^2,
\end{equation}
where we introduced a gap to QCP as $\Delta_{\text{QCP}}(\Gamma)=\sqrt{\gamma(\Gamma)}\Omega_c$. The most striking feature of this result is that proliferation of quantum fluctuations to finite temperatures gives a power law instead of exponential behavior even in the presence of a full SC gap. As is known, a power law in the specific heat occurs only in the unconventional superconductors having nodal structure of the gap. In particular $\propto T^2$ is a characteristic of a gap structure with first-order nodes at isolated points. 

We note that the details of the microscopic model enter Eq. \eqref{C-QF} only via the ratio $v_F Q_c/\Omega_c$ so that $T^2$ dependence is a model independent result.
Furthermore, as $\Gamma_c/2\pi T_{c0}\ll1$ for a broad range of parameters, then to a good approximation $v_F Q_c = \Delta \sqrt{\pi}$ and $\Omega_c = \Delta \sqrt{\pi/2}$ leading to a universal expression, $\delta C_{\text{QF}}=(18\zeta(3)/\pi)(T/v_F)^2$.
Ultimately, at the lowest temperatures, $T<\Delta_{\text{QCP}}$, specific heat crosses over to exponential dependence, $\delta C_{\text{QF}}\propto (\Delta_{\text{QCP}}/T) e^{-\Delta_{\text{QCP}}/T}$. 
We note that the same conclusion has been reached independently in the considerations of a different model \cite{Carvalho}. 

\begin{figure*}[t!]
  \centering
  \includegraphics[width=7in]{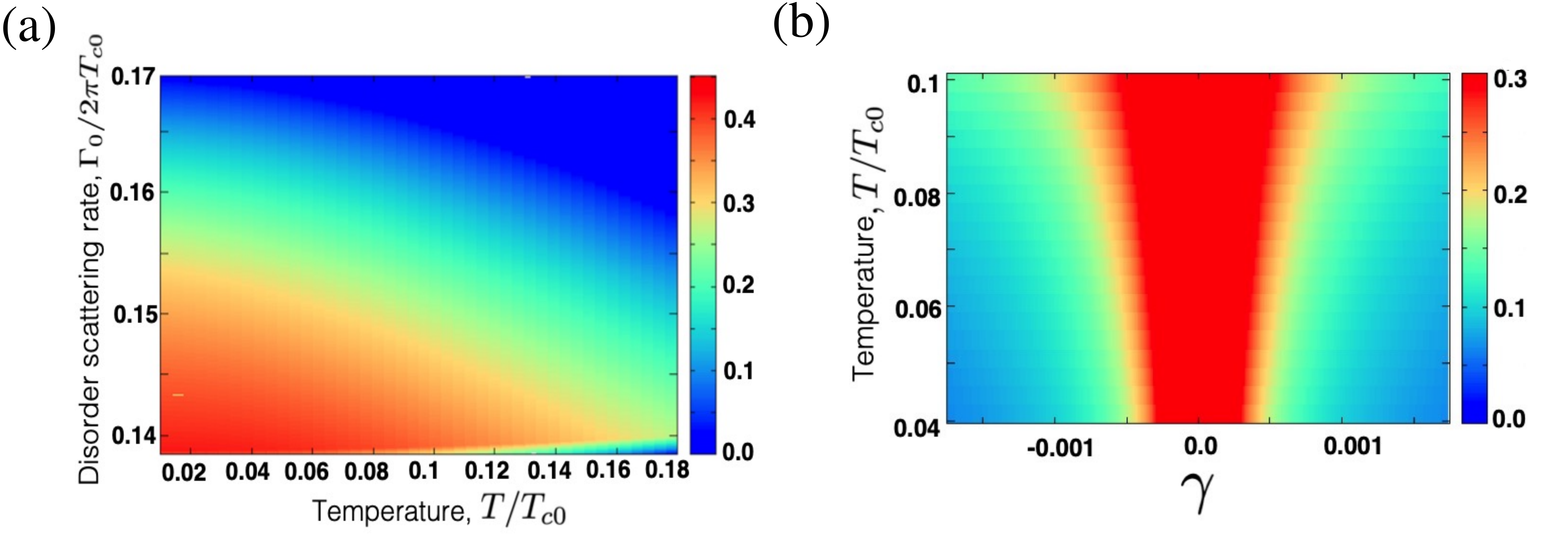}
\caption{(a) Contour plot of the London penetration depth
$\lambda^{-2}(\Gamma,T)$ (arb. units) calculated within the MF theory approximation as a function of temperature and disorder scattering rate $\Gamma_0$ assuming $\Gamma_\pi=0.4\Gamma_0$. We note that already at the MF the width of the region in which  $\lambda^{-2}$ has maximum value narrows upon an increase in temperature. (b) Normalized quantum fluctuation correction to the electromagnetic response kernel [Eq. \eqref{K-SDW}] as a function of the proximity to the QCP gap $\gamma$, showing an emergent peak in a color plot.}
  \label{lambda-3D-Visual}
\end{figure*}

\subsection{Penetration depth near QCP}

We turn our attention to the anomalies in the magnetic penetration depth $\lambda(T,x)$, where numerous recent measurements \cite{Matsuda,Auslaender-1,Auslaender-2,Zheng,Prozorov} revealed a distinct peak in the low-temperature limit $T\ll\Delta$ concentrated around the putative QCP $x\to x_c$. The model we explore in this study with $x=\Gamma$ is perhaps best suited to experiments of Ref. \cite{Prozorov} on Ba(Fe$_{1-x}$Co$_x$)$_2$As$_2$. This material is in the disordered limit with fully gapped Fermi surface as opposed to BaFe$_2$(As$_{1-x}$P$_x$)$_2$, which is a rather clean system that displays nodal superconductivity. However, the arguments we put forward are rather generic and in fact apply to both compounds.   

It is natural to account for soft bosonic modes in the fermionic electromagnetic response function that defines $\lambda(x,T)$. However, the one-loop fluctuation correction while giving a good approximation outside the critical region is inapplicable inside this region.
In the present context, this implies that as a matter of principle, the character of the $\lambda(x)$ singularity cannot be determined on the level of a one-loop approximation. Indeed, the mean field theory predicts a deep in $\lambda(x)$ \cite{Dzero,Fernandes-London,Kuzmanovski-London}, see also Fig. \ref{lambda-3D-Visual} for the further illustration. Therefore, in order to turn the deep into a peak the fluctuation correction must exceed the mean field value. According to the Ginzburg criterion, however, this cannot happen in the region of validity of one-loop approximation. For this reason, the problem has to be solved inside a critical region and is essentially nonperturbative.

Such a solution valid in critical region is possible at $T=0$ for the model of electrons coupled to critical bosons with the mass term $\propto (x- x_c)$ \cite{Chowdhury-PRL}. In this model there is a universal relation between the critical scaling of $\lambda$ at $x$ above and below $x_c$. When the bosons are viewed as collective fermion excitations as captured by $L_{Q,\Omega_m}$ such a universal relation is lost as the ratio of the paramagnon masses at $x = x_c \pm \delta$ is a model dependent number, while in, e.g., Ising boson theory it is $2$. In our specific model this number $\gamma'_+/\gamma'_-$ is a nonuniversal function of $\Gamma_{0,\pi}$. Despite this discrepancy with the purely bosonic approach, the $x$ dependence of $\lambda$ established in Ref.~\cite{Chowdhury-PRL}  remains monotonic in our model as well.
This leaves us with the unresolved puzzle of the peak in $\lambda(x)$.

Our solution to this problem builds on the strong $x$ dependence of the $T$-dependent part of the penetration depth, $\lambda(T)-\lambda(T=0)$. 
In distinction with the $T=0$ case, at the mean field level $\delta\lambda(T)=\lambda(T) - \lambda(0)\propto e^{-\Delta/T}$ is suppressed exponentially at $T\ll\Delta$. 
Therefore, just outside the critical region the one-loop correction gives a reliable estimate of fluctuation correction to $\delta\lambda(T)$. 
This correction yields the peak in $\lambda(T)$ at the temperatures 
$T\gtrsim \Delta^2/E_F$ much smaller than $\Delta$.

To quantify these statements we express the fluctuation correction to $\lambda=\lambda_0+\delta\lambda_{\text{QF}}$ through the correction to the static, long wave length limit of the current correlation function $K=K_0+\delta K_{\text{QF}}$ (see Appendix \ref{App:Fluctuations}):
\begin{equation}
\frac{\delta\lambda_{\text{QF}}}{\lambda_0}=-\frac{\delta K_{\text{QF}}}{2K_0}, \quad K_0=\frac{1}{2}\nu_Fe^2v^2_F,\quad \lambda^{-2}_{0}=\frac{4\pi}{w} K_0, 
\end{equation}
where $w$ is the interlayer separation as appropriate to the quasi-2D systems.
The one-loop correction of the electromagnetic kernel $K_{\text{QF}}$ contains both effective mass renormalization, captured by the density of states (DOS) type diagrams, and vertex renormalization  expressed by the Maki-Thompson (MT) type quantum interference processes. The Aslamazov-Larkin vertex corrections cancel for the case when the gaps on hole and electron Fermi surfaces are of equal magnitude (and opposite sign), which is implicit in our model. The cancellation occurs at the level of fermionic triangular blocks as for each block there are two ways to arrange electron and hole Green's function lines and their corresponding momenta which cancel each other. We thus have
\begin{equation}\label{dgamma}
\partial_{\gamma}\delta\! K_{\text{QF}}\!=\!\frac{\mathcal{N}}{2}\!e^2 v_F^2 \Tr\left[\partial_\gamma L_{Q,\Omega_m}\right]F_{\text{l}},\, \, F_{\text{l}}=F_{\text{DOS}}\!+\!F_{\text{MT}}.   
\end{equation}
Apart from excluding transverse spin excitations, taking the derivative of $\delta K_{\text{QF}}$ makes the integration over the boson energies and momenta convergent at the ultraviolet.
This means that at $\gamma \ll 1$ the important values of $Q$ and $\Omega$ are within the region of applicability of low-energy expansion of $L_{Q,\Omega_m}$ assumed above. 
At the same time the integrations over fermion and boson energies and momenta in Eq.~\eqref{dgamma} factorize,
and the fermionic loop $F_{\text{l}}$ can be taken at zero boson energy and momentum $(Q,\Omega_m)\to0$. 
The factorization in Eq.~\eqref{dgamma} is possible thanks to the energy scale separation of fermions $\Delta$ and bosons $\Delta_{\text{QCP}}\ll \Delta$.  
The individual terms are $F_{\text{DOS}}=2 \Tr\left[V_S^2 \hat{G}\hat{\tau}_3 \hat{G}\hat{\Xi}^z \hat{G} \hat{\Xi}^z \hat{G}\hat{\tau}_3\right]$ and $F_{\text{MT}}=\Tr\left[V_S^2 \hat{G}\hat{\tau}_3 \hat{G}\hat{\Xi}^z \hat{G}
\hat{\tau}_3 \hat{G} \hat{\Xi}^z\right]$,
where the spin vertex renormalization $V_S$ can be evaluated at $(Q,\Omega_m)\to0$ taken in any order due to the non-conservation of the magnetization (see Appendix \ref{App:Disorder} for exhaustive details). 
For $\Gamma_{\pi}=0$, $V_S = (1 + \Gamma_0/\sqrt{\Delta^2 + \omega^2_n})^{-1}$, where $\omega_n$ is a frequency argument of Green functions. In the wide range of parameters $(T,\Gamma_c) \ll \Delta$, $F_{\text{l}}\simeq \nu_F/\Delta^2$.

We further separate $\delta K_{\text{QF}}=\delta K_{\text{QCP}}+\delta K_{\text{SDW}}$ into zero-temperature $(\delta K_{\text{QCP}})$ and finite-temperature $(\delta K_{\text{SDW}})$ terms. For the former we straightforwardly find in a limit, $\Gamma_{c} \ll \Delta$ 
\begin{equation}\label{K-QCP}
\frac{\delta K_{\text{QCP}}}{K_0}=- \sqrt{\frac{\pi}{2}}\frac{3 \pi \mathcal{N}}{16}  \frac{\Delta}{E_F}\sqrt{\gamma} 
\end{equation}
up to a constant with the Fermi energy, $E_F = \pi \nu_F v_F^2/4$.
This result applies at both sides of QCP, and complements a similar calculation in a paramagnetic phase done in a band model of the QCP \cite{LVKC}.
This result gives positive correction to the penetration depth, however as we discussed above, is insufficient to turn it into a peak within the validity of perturbative analysis. 
To elucidate this point, we introduce the Ginzburg parameter $\mathrm{Gi}$ given by a $\sqrt{\gamma}$ such that the fluctuation correction, $\partial_{\gamma}\delta K_{\text{QCP}}$ becomes comparable to the mean field value. It follows that the loop expansion is a series in powers of $\mathrm{Gi}/\gamma$. For instance, the two-loop contribution can be estimated to give a correction to $\delta K_{\text{QCP}}/K_0 $ of the form, $\propto\mathrm{Gi}^2\gamma^{-1/2}$ (see Appendix \ref{App:Fluctuations} for extensive details). From the comparison to Eq. \eqref{K-QCP} we then conclude that $\mathrm{Gi}=\Delta/E_F$.

We proceed to analyze the temperature dependent part of the response kernel. After the Matsubara sum we arrive at 
\begin{equation}\label{K-SDW}
\frac{\partial_\gamma\delta K_{\text{SDW}}}{K_0}=-\frac{\mathcal{N}}{2\pi}\frac{\Omega^4_c F_{\text{l}}}{\nu^2_F}\int_Q\frac{f(E_Q/2T)}{E^3_Q},
\end{equation}
where $E_Q=\Omega_c\sqrt{\gamma+(Q/Q_c)^2}$ and $f(z)=\coth(z)-1+z/\sinh^2(z)$. In the temperature range above the QCP gap, $\Delta_{\text{QCP}}<T<\Delta$, we find for the penetration depth 
\begin{equation}\label{lambda-QF}
\frac{\delta\lambda_{\text{QF}}}{\lambda_0}=\frac{\mathcal{N}}{8}\frac{T}{E_F}\ln\left(\frac{1}{\gamma}\right),
\end{equation}
so that the peak height is estimated as  $\delta\lambda^{\text{max}}_{\text{QF}}/\lambda_0\simeq(T/\Delta)\text{Gi}\ln(1/\text{Gi})$. At temperatures within the QCP gap, $T<\Delta_{\text{QCP}}$, we instead find an exponential dependence $\delta\lambda_{\text{QF}}/\lambda_0\propto e^{-\Delta_{\text{QCP}}/T}$. 
The linear in $T$ result holds in both paramagnetic and magnetically ordered phases.
The only difference originates from the difference in the coefficients $\gamma'_{\pm}$ describing the paramagnon softening in the two phases as introduced above. 

\section{Discussion, summary, and outlook}\label{Sec:Summary}

To address implications of these results in light of experiments we stress that measurements of Ref. \cite{Prozorov} on Ba(Fe$_{1-x}$Co$_x$)$_2$As$_2$ were carried out at $T\sim4.5$K (with maximal $T_c\sim 25$K), whereas measurements of Ref. \cite{Matsuda} on BaFe$_2$(As$_{1-x}$P$_x$)$_2$ were done at $T=1.2$K (with maximal $T_c\sim 30$K). In the Co-doped case we interpret the emergence of the peak as due to SDW fluctuations at finite temperature once the system is tuned into the anomalous region by doping so that $\lambda_{\text{QF}}$ from Eq. \eqref{lambda-QF} dominates over suppressed mean field behavior $\delta\lambda_{\text{MF}}\propto e^{-\Delta/T}$. This is exemplified in Fig. \ref{Fig-QCP-PD}(b) and further in Fig. \ref{lambda-3D-Visual}(b). In addition, due to renormalization of fluctuations by finite $M$ in the phase of coexistence the structure of the peak should be nonsymmetric from both sides of QCP. This is supported by our model analysis and is in qualitative agreement with observations. 

In contrast, in the P-doped case a quasi-linear-$T$ dependence of $\lambda$ was seen and attributed to the nodal structure of the gap. However, it is crucial to point out that the slope of this linear behavior was changing with doping attaining a maximum at QCP (see Fig. 3 of Ref. \cite{Matsuda}). We attribute this enhancement to SDW fluctuations which also result in linear-in-$T$ penetration depth as we show in Eq. \eqref{lambda-QF}. Indeed, this statement can be made more precise as in the case of a SC with simple isolated nodes, Eq. \eqref{lambda-QF} defines the renormalization of the relative slope in $T$-linear behavior of the penetration depth, $\delta\lambda(T)/\lambda_0=s(T/\Delta)$, so that slope receives a correction $\delta s\simeq \text{Gi}\ln |x-x_c|^{-1}$, which becomes progressively steeper towards a QCP. This prediction is thus consistent with corresponding behavior seen in experiments of Ref. \cite{Matsuda}.  

A signature of the peak was also detected in (Ba$_{1-x}$K$_x$)Fe$_2$As$_2$ \cite{Auslaender-2}
concomitant with nonmonotonic doping dependence and change in $\delta \lambda\propto T^n$ power law \cite{Cho}. While SDW fluctuations certainly play an important role near QCP, interpretation of the data in the whole range is difficult as K-doped system displays a series of Lifshitz topological phase transitions resulting in gaped-to-nodal change of the pairing gap. Additional complications come from the apparent narrow dome of $s+is'$ superconductivity separating gapped and nodal regions \cite{Klauss} capturing which is beyond our two-band model.  

In summary, in this work we studied the interplay of magnetism and superconductivity in the context of iron pnictides. Our principal results are Eqs. \eqref{C-QF} and \eqref{lambda-QF} for the temperature dependence of the specific heat and the London penetration depth, respectively, due to physics associated with the QCP. These results are significant as power-law $T$ dependence of thermodynamic properties of SCs is used as a hallmark diagnostic for their unconventional character, namely determination of the types of the nodes of superconducting order parameter. Yet we demonstrate that even in the presence of the full gap such behavior can be promoted by the quantum criticality under the dome of superconductivity.     

We further comment that there remain some unresolved issues that warrant additional studies. In particular, a double-peak structure was detected in the penetration depth measurements in NaFe$_{1-x}$Co$_x$As \cite{Zheng}. This remarkable feature was attributed to the second putative QCP of nematic origin.  However, the mere statement of multiple possible QCPs under the SC dome is at odds with the present state of the theory \cite{Fernandes-QCP} that predicts that magnetic and nematic transitions merge together into the weak first-order quantum critical line that thus terminates at a single QCP.   

In closing, we mention that our results open interesting perspectives for the studies of transport properties due to QCP, specifically for the optical conductivity and thermoelectric effects, where one may hope to obtain anomalous frequency and temperature dependencies due to quantum fluctuations. It is also of special interest to investigate the QCP behavior due to the interplay of charge/pair-density order and superconductivity, which is highly relevant topic in the physics of cuprates.     

\section*{Acknowledgments} 

We thank E. Berg, V. S. de Carvalho, A. Chubukov, R. Fernandes, S. Gazit, Y. Matsuda, D. Orgad, R. Prozorov, S. Sachdev, J. Schmalian, and T. Shibauchi for important discussions that shaped this study. This work was supported in part by BSF Grant No. 2016317, ISF Grant No. 2665/20 (M.K.), NSF-DMR-BSF-2002795 (M.D. and M.K.), and U. S. Department of Energy (DOE), Office of Science, Basic Energy Sciences (BES) Program for Materials and Chemistry Research in Quantum Information Science under Award No. DE-SC0020313 (A.L.) and DE-SC0016481 (M.D.). This work was performed in part at the Landau Institute for Theoretical Physics, Max Planck Institute for the Physics of Complex Systems, and at the Aspen Center for Physics, which is supported by National Science Foundation Grant No. PHY-1607611.

\appendix

\section{Quasiclassical theory}\label{App:QuasiClass}

The purpose of this section is to provide the extended details on the quasiclassical approximation in the context of SC-SDW coexistence and calculate disorder-averaged single particle propagator in the framework of the Eilenberger equation. The results of this section are primarily based on the prior analysis of Refs. \cite{Vavilov,Dzero,Kirmani}. 

\subsection{Hubbard-Stratonovich transformation}

The Hamiltonian of the model we consider in Eq. \eqref{Eq:H} contains two four-fermion terms corresponding to magnetic [Eq. \eqref{Eq:Hsdw}] and superconducting [Eq. \eqref{Eq:Hsc}] interactions. The corresponding mean-field Hamiltonian, $\hat{H}=\sum_{\bm{p}}\Psi^\dag_{\bm{p}}\hat{\mathcal{H}}_{\bm{p}}\Psi_{\bm{p}}$, with $\hat{\mathcal{H}}_{\bm{p}}=\hat{\mathcal{H}}_0+\hat{\mathcal{H}}_{\textrm{mf}}$ and $\hat{\mathcal{H}}_0=\xi_{\bm{p}}\hat{\tau}_3\hat{\rho}_3\hat{\sigma}_0$, which is bilinear in fermion $\Psi_{\bm{p}}$ operators, can be obtained by decoupling interaction terms. This is achieved in a standard way with the Gaussian integral of Hubbard-Stratonovich transformation that invokes two additional fields $\Delta$ and $M$ associated with the superconducting and magnetic order parameters. For the sign-changing $s^{\pm}$ pairing we will have
\begin{equation}\label{AnomFs}
g_{\text{sc}} \langle \hat{c}_{\bm{p}\up}\dg\hat{c}_{-\bm{p}\dn}\dg\rangle=-|\Delta|e^{i\phi},\quad 
g_{\text{sc}}\langle \hat{f}_{\bm{p}\up}\dg\hat{f}_{-\bm{p}\dn}\dg\rangle=|\Delta|e^{i\phi}. 
\end{equation}
Furthermore, we assume that the direction of magnetization is fixed along the $z$ axis. According to this convention after integration we arrive at 
\begin{equation}\label{HmfSc}
\hat{\mathcal{H}}_{\textrm{mf}}=-|\Delta|\left[\cos\phi(\hat{\tau}_3\hat{\rho}_1\hat{\sigma}_0)+\sin\phi(\hat{\tau}_3\hat{\rho}_2\hat{\sigma}_0)\right]+M\hat{\tau}_1\hat{\rho}_0\hat\sigma_3.
\end{equation}
In what follows, and without loss of generality, we consider the superconducting order parameter to be real, thus setting $\phi=0$.

\subsection{Self-energy in Dyson equation}

In order to determine the disorder averaged single-particle Green's function $\hat{G}_{\bm{p},\omega_n}$ we must solve the matrix Dyson equation  
\begin{equation}\label{Gpwn}
\left[i\omega_n-\hat{\mathcal{H}}_{\bm{p}}-\hat{\Sigma}_{\omega_n}\right]\hat{G}_{\bm{p},\omega_n}=\hat{1}.
\end{equation}
Here $\hat{\Sigma}_{\omega_n}$ is the frequency dependent self-energy, which is generated by the disorder. In this work we consider the spin-independent (i.e. nonmagnetic) disorder potential defined by Eq. \eqref{Eq:Hdis}, which however includes interband transitions thus it is nondiagonal in the band basis. Within the self-consistent Born approximation, the corresponding expression for the self-energy reads
\begin{align}\label{Sigma}
\hat{\Sigma}_{\omega_n}=\frac{\Gamma_0}{\pi\nu_F}\int\frac{d^2\bm{p}}{(2\pi)^2}(\hat{\tau}_0\hat{\rho}_3\hat{\sigma}_0)
\hat{G}_{\bm{p},\omega_n}(\hat{\tau}_0\hat{\rho}_3\hat{\sigma}_0)\nonumber \\ 
+\frac{\Gamma_\pi}{\pi\nu_F}\int\frac{d^2\bm{p}}{(2\pi)^2}(\hat{\tau}_1\hat{\rho}_3\hat{\sigma}_0)
\hat{G}_{\bm{p},\omega_n}(\hat{\tau}_1\hat{\rho}_3\hat{\sigma}_0),
\end{align}
where the cross terms vanish and $\Gamma_{0,\pi}\propto \nu_F|U_{0,\pi}|^2$ are the corresponding disorder scattering rates.

\subsection{Single particle propagator}

The self-energy can be computed within the quasiclassical approximation. For this purpose, consider the following reduced Eilenberger function
\begin{equation}\label{EilenGwn}
\hat{\cal G}_{\omega_n}=\frac{i}{\pi\nu_F}\int\frac{d^2\bm{p}}{(2\pi)^2}(\hat{\tau}_3\hat{\rho}_3\hat{\sigma}_0)\hat{G}_{\bm{p},\omega_n}. 
\end{equation}
The self-energy part can now be expressed in terms of this function
\begin{align}\label{Sigma4G}
&\hat{\Sigma}_{\omega_n}=-i\Gamma_{0}(\hat{\tau}_0\hat{\rho}_3\hat{\sigma}_0\hat{\tau}_3\hat{\rho}_3\hat{\sigma}_0)
\hat{\cal G}_{\omega_n}(\hat{\tau}_0\hat{\rho}_3\hat{\sigma}_0)\nonumber \\ 
&-i\Gamma_{\pi}(\hat{\tau}_1\hat{\rho}_3\hat{\sigma}_0
\hat{\tau}_3\hat{\rho}_3\hat{\sigma}_0)\hat{\cal G}_{\omega_n}(\hat{\tau}_1\hat{\rho}_3\hat{\sigma}_0).
\end{align}
The quasiclassical function $\hat{\cal G}_{\omega_n}$ can be found from the self-consistent solution of the Eilenberger-Dyson equation 
\begin{equation}\label{Eq1Eilen}
\left[i\omega_n\hat{\tau}_3\hat{\rho}_3\hat{\sigma}_0;\hat{\cal G}\right]-\left[\hat{\mathcal{H}}_{\textrm{mf}}\hat{\tau}_3\hat{\rho}_3\hat{\sigma}_0;\hat{\cal G}\right]
-\left[\hat{\Sigma}_{\omega}\hat{\tau}_3\hat{\rho}_3\hat{\sigma}_0;\hat{\cal G}\right]=0, 
\end{equation}
where $[\hat{A};\hat{B}]$ represents a commutator of two matrices in each term, respectively. To resolve this matrix equation we need a proper parametrization. The particularly convenient ansatz for the quasiclassical function reads
\begin{equation}\label{Anzats}
\hat{\cal G}_{\omega_n}=g_{\omega_n}(\hat{\tau}_3\hat{\rho}_3\hat{\sigma}_0)
+if_{\omega_n}(\hat{\tau}_0\hat{\rho}_2\hat{\sigma}_0)+is_{\omega_n}(\hat{\tau}_2\hat{\rho}_3\hat{\sigma}_3),
\end{equation}
and contains now in addition to normal-$g_{\omega_n}$ and anomalous-$f_{\omega_n}$ also a magnetic-$s_{\omega_n}$ Green's function.  
The expression for the self-energy can be then further reduced to the form 
\begin{equation}\label{SigmaSCBA}
\hat{\Sigma}_{\omega_n}\!\!=-i\Gamma_{\text{t}}g_{\omega_n}(\hat{\tau}_0\hat{\rho}_0\hat{\sigma}_0)+i\Gamma_{\text{m}}f_{\omega_n}
(\hat{\tau}_3\hat{\rho}_1\hat{\sigma}_0)-i\Gamma_{\text{t}}s_{\omega_n}(\hat{\tau}_1\hat{\rho}_0\hat{\sigma}_3),
\end{equation}
where we introduced total $(\Gamma_{\text{t})}$ and effectively pair-breaking ($\Gamma_{\text{m}}$) scattering rates 
\begin{equation}\label{Eq:Gamma-tm}
\Gamma_{\text{t,m}}=\Gamma_0\pm\Gamma_\pi.
\end{equation}
Now, we use these expressions to evaluate the commutators in 
Eq. (\ref{Eq1Eilen}):
\begin{subequations}\label{Comms}
\begin{align}
\left[i\omega_n\hat{\tau}_3\hat{\rho}_3\hat{\sigma}_0;\hat{\cal G}\right]&=2i\omega_n\left(f_{\omega_n}\hat{\tau}_3\hat{\rho}_1\hat{\sigma}_0+s_{\omega_n}\hat{\tau}_1\hat{\rho}_0\hat{\sigma}_3\right), \\
\left[\hat{\mathcal{H}}_{\textrm{mf}}\hat{\tau}_3\hat{\rho}_3\hat{\sigma}_;\hat{\cal G}\right]&=-2g_{\omega_n}\left(\Delta\hat{\tau}_3\hat{\rho}_1\hat{\sigma}_0-M\hat{\tau}_1\hat{\rho}_0\hat{\sigma}_3 \right)\nonumber \\ &-2i(s_{\omega_n}\Delta+f_{\omega_n}M)
{\hat{\tau}_2\hat{\rho}_1\hat{\sigma}_3}, \\
\left[\hat{\Sigma}_{\omega}\hat{\tau}_3\hat{\rho}_3\hat{\sigma}_0;\hat{\cal G}\right]&=-4i\Gamma_\pi g_{\omega_n}f_{\omega_n}
\hat{\tau}_3\hat{\rho}_1\hat{\sigma}_0\nonumber \\ &-4i\Gamma_{\text{t}} g_{\omega_n}s_{\omega_n}
\hat{\tau}_1\hat{\rho}_0\hat{\sigma}_3-4\Gamma_0 s_{\omega_n}f_{\omega_n}
{\hat{\tau}_2\hat{\rho}_1\hat{\sigma}_3}.
\end{align}
\end{subequations}
Next we collect the coefficients in front of each matrix appearing in (\ref{Comms}) and thus obtain a closed set of coupled algebraic equations   
for yet unknown functions $g_{\omega_n}$, $f_{\omega_n}$, and $s_{\omega_n}$
\begin{subequations}\label{BulkMF}
\begin{align}
&\left(\omega_n+2\Gamma_\pi g_{\omega_n}\right)f_{\omega_n}=i\Delta g_{\omega_n} , \\ 
&\left(\omega_n+2\Gamma_{\text{t}}g_{\omega_n}\right)s_{\omega_n}=-iM g_{\omega_n}, \\ 
&\Delta s_{\omega_n}+M f_{\omega_n}=2i\Gamma_0 f_{\omega_n}s_{\omega_n},
\end{align}
\end{subequations}
which should be also supplemented by the normalization condition:
\begin{equation}\label{norm}
g_{\omega_n}^2-f_{\omega_n}^2-s_{\omega_n}^2=1.
\end{equation}
Note that the last equation (\ref{BulkMF}) is redundant and, importantly, functions $f_{\omega_n}$, $s_{\omega_n}$ are purely imaginary.
If we now insert Eq. (\ref{SigmaSCBA}) into Eq. (\ref{Gpwn}), we can group the terms which have identical matrix structure together. Introducing then 
\begin{subequations}\label{tildas}
\begin{align}
\varpi_n=\omega_n+\Gamma_{\text{t}}g_{\omega_n}, \\ 
\Delta_{\omega}=\Delta-i\Gamma_{\text{m}}f_{\omega_n}, \\
 M_{\omega}=M-i\Gamma_{\text{t}}s_{\omega_n}, 
\end{align}
\end{subequations}
we find for the single particle propagator 
\begin{align}\label{GpwnDis}
&\hat{G}_{\bm{p},\omega_n}=\frac{1}{\xi_{\bm{p}}^2+D_{\omega}^2+M_{\omega}^2}\nonumber\\
&\times\left[-i\varpi_n\hat{\tau}_0\hat{\rho}_0\hat{\sigma}_0+\xi_{\bm{p}}\hat{\tau}_3\hat{\rho}_3\hat{\sigma}_0+\Delta_\omega\hat{\tau}_3\hat{\rho}_1\hat{\sigma}_0-M_\omega\hat{\tau}_1\hat{\rho}_0\hat{\sigma}_3\right], 
\end{align}
where for the future use we introduced a notation 
\begin{equation}\label{Eq:D}
D_\omega=\sqrt{\Delta_\omega^2+\varpi_n^2}. 
\end{equation}
Equation \eqref{GpwnDis} is the main result of this section.


\begin{figure}
    \centering
    \includegraphics[width=2.5in]{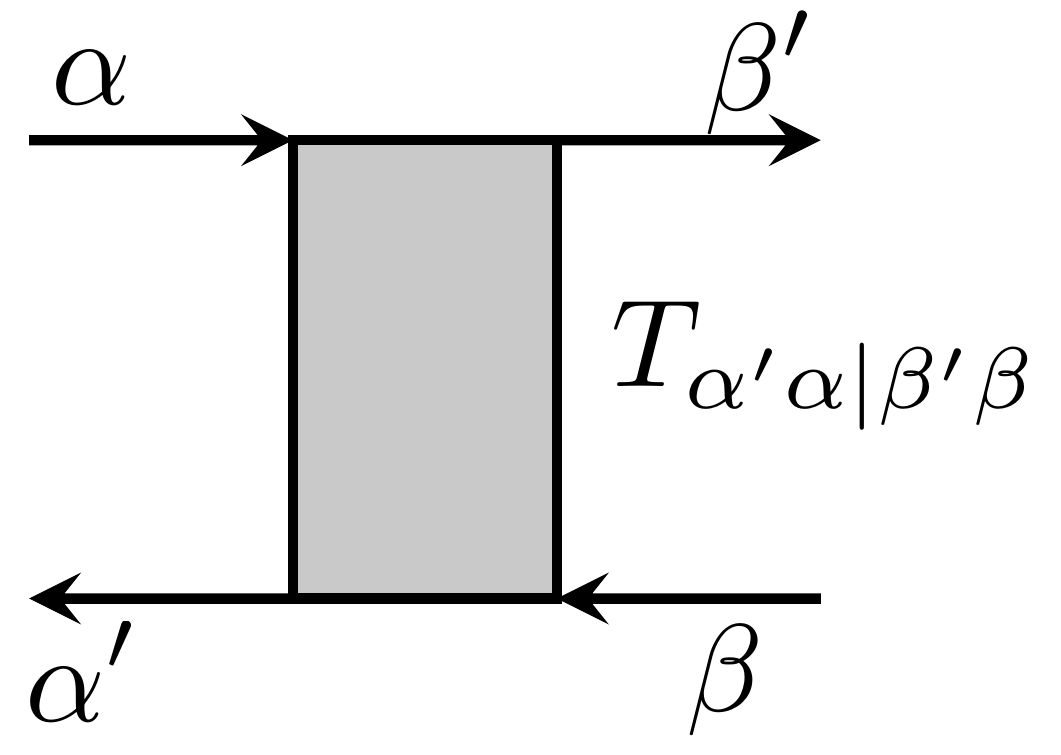}
    \caption{In the figure $T$ represents the diagrammatic block such as the disorder vertical line.
The incoming indices $\alpha$, $\beta$,  and outgoing indices $\alpha'$, $\beta'$ all operate in the eight-dimensional space of the direct product of fermion pockets, Nambu, and spin sub spaces. It can be written as the sequence of the corresponding three indices $\alpha \equiv (\alpha_{\tau},\alpha_{\rho},\alpha_{\sigma})$, where each of the three indices assumes two values. In this way the block $\hat{T}$ becomes a $64 \times 64$ matrix.}
    \label{Fig:T-matrix}
\end{figure}

\section{Disorder dressed vertex in a superbasis}\label{App:Disorder}

In order to study the dressing by disorder we use the following representation of the diagrammatic block, such as an impurity line, depicted 
in Fig. \eqref{Fig:T-matrix}
\begin{align}\label{T-dis}
T_{\alpha'\alpha|\beta'\beta}= \sum_{\stackrel{\mu\nu\gamma}{\mu'\nu'\gamma'}}  \Gamma_{\mu\nu\gamma|\mu'\nu'\gamma'} [\hat{\tau}_{\mu} \hat{\rho}_{\nu} \hat{\sigma}_{\gamma}]_{\alpha'\alpha} [\hat{\tau}_{\mu'} \hat{\rho}_{\nu'} \hat{\sigma}_{\gamma'}]_{\beta'\beta}. 
\end{align}
Here each label such as $\alpha$ in fact contains three labels $\alpha = (\alpha_{\tau},\alpha_{\rho},\alpha_{\sigma})$, where each of the labels $\alpha_{\tau}$, $\alpha_{\rho}$, and $\alpha_{\sigma}$ runs over two possible values $\pm 1$, so that in total the label $\alpha$ runs over eight values. The pair of indices $\alpha\alpha'$ can have 64 independent values. The indices $\mu$, $\nu$, and $\gamma$ run over four indices, $0,1,2,3$. The total number of such indices $4^3 = 64$ is sufficient to parametrize all the combinations.
The same counting  holds for the second pair of indices $\beta\beta'$ parametrized by the second triplet of indices $\mu'\nu'\gamma'$.
Another way to convince oneself that the representation in Eq. \eqref{T-dis} is always possible for any $T_{\alpha'\alpha|\beta'\beta}$ is to count the number of free parameters on the right and on the left hand side of Eq. \eqref{T-dis}. In both cases we have $(2^3)^4 = (4^3)^2$.

The coefficients in the decomposition of Eq. \eqref{T-dis} are given by
\begin{equation}\label{dec13}
\Gamma_{\mu\nu\gamma|\mu'\nu'\gamma'} = \frac{1}{8^2} \sum_{\stackrel{\alpha\alpha'}{\beta\beta'}}
T_{\alpha'\alpha|\beta'\beta}  [\hat{\tau}_{\mu} \hat{\rho}_{\nu} \hat{\sigma}_{\gamma}]_{\alpha\alpha'} [\hat{\tau}_{\mu'} \hat{\rho}_{\nu'} \hat{\sigma}_{\gamma'}]_{\beta\beta'}. 
\end{equation}
For the two types of the disorder in the model, we thus have for the block structure 
\begin{align}\label{dec15}
&T_{\alpha'\alpha|\beta'\beta}  = T^0_{\alpha'\alpha|\beta'\beta}  + T^{\pi}_{\alpha'\alpha|\beta'\beta}, \nonumber \\ 
&T^{0}_{\alpha'\alpha|\beta'\beta} = u_0^2 [\hat{\tau}_0\hat{\rho}_3\hat{\sigma}_0]_{\beta'\alpha}[\hat{\tau}_0\hat{\rho}_3\hat{\sigma}_0]_{\alpha'\beta},\nonumber\\  
&T^{\pi}_{\alpha'\alpha|\beta'\beta} =u_{\pi}^2 [\hat{\tau}_1\hat{\rho}_3 \hat{\sigma}_0]_{\beta'\alpha} [\hat{\tau}_1\hat{\rho}_3 \hat{\sigma}_0]_{\alpha'\beta}
\end{align}
Correspondingly, Eq. \eqref{dec13} gives, using the reduced expressions for the rates, $u_0^2 = (\pi \nu_F)^{-1} \Gamma_0$ and $u_{\pi}^2 = (\pi \nu_F)^{-1} \Gamma_{\pi}$, for the total matrix in the expansion
\begin{subequations}\label{dec17a}
\begin{align}
[\hat{\Gamma}]_{\mu\nu\gamma|\mu'\nu'\gamma'} &= [\hat{\Gamma}_\pi]_{\mu\nu\gamma|\mu'\nu'\gamma'} + [\hat{\Gamma}_0]_{\mu\nu\gamma|\mu'\nu'\gamma'}, \\
[\hat{\Gamma}_0]_{\mu\nu\gamma|\mu'\nu'\gamma'} &= \frac{u^2_0}{8} \delta_{\mu\mu'}\delta_{\nu\nu'} \delta_{\gamma\gamma'}[ \delta_{\nu,0} + \delta_{\nu,3} -  \delta_{\nu,2} - \delta_{\nu,1}],\\
[\hat{\Gamma}_\pi]_{\mu\nu\gamma|\mu'\nu'\gamma'}  &= \frac{u^2_\pi}{8} \delta_{\mu\mu'}\delta_{\nu\nu'} \delta_{\gamma\gamma'}
[ \delta_{\mu,0} + \delta_{\mu,1} -  \delta_{\mu,2} - \delta_{\mu,3}]\nonumber\\ 
 &\times[ \delta_{\nu,0} + \delta_{\nu,3} -  \delta_{\nu,2} - \delta_{\nu,1}].
\end{align}
\end{subequations}

The key advantage of the representation defined by Eq. \eqref{T-dis} is that it allows us to turn the disorder vertical lines insertions as the horizontal blocks amenable to regular geometrical series summation in analogy with the random phase approximation (RPA). The conceptual complication here is that, in the present case, we have a geometrical series of $64 \times 64$ matrices, nevertheless the formal expansions become similar to the usual RPA as we demonstrate below.

\begin{figure}
    \centering
    \includegraphics[width=3.25in]{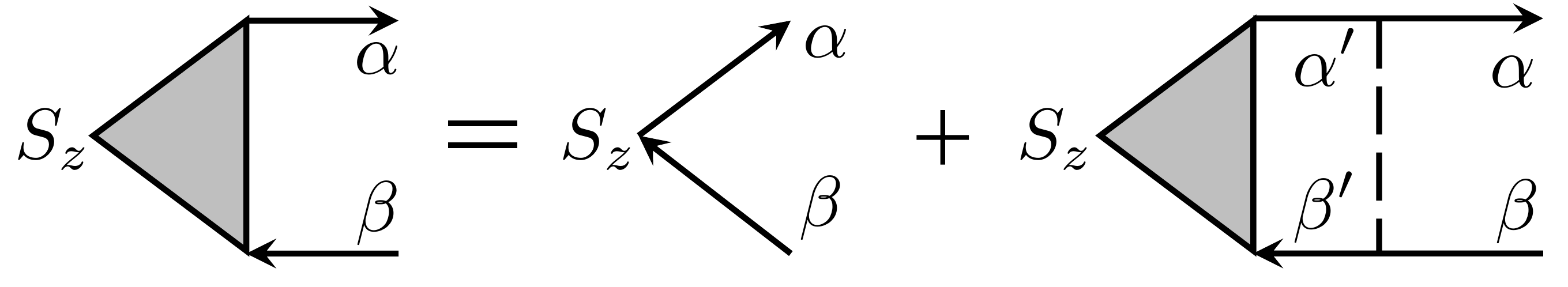}
    \caption{The diagrammatic representation of Eq. \eqref{Ladder_Sz} for the spin vertex renormalization by disorder computed in the ladder approximation. A single impurity is depicted by a dashed line whereas the shaded region represents the whole geometric series of impurity lines ladder. Again the indices $\alpha$ and $\beta$ stand for the triplet of indices, namely, $\alpha \equiv (\alpha_{\tau},\alpha_{\rho},\alpha_{\sigma})$ and $\beta \equiv (\beta_{\tau},\beta{\rho},\beta_{\sigma})$. This form makes it possible to apply the standard RPA resummation.}
    \label{Fig:Sz-Ladder}
\end{figure}

Let us illustrate now how this construction works in practice. For this purpose, we consider dressing of the spin vertex by the disorder ladder. 
We carry out this computation with two simplifying observations that can be explicitly verified \textit{a posteriori}. (i) We focus on a nonmagnetic case, $M=0$, since finite $M$ only introduces a regular in $M$ corrections that vanish at the QCP. (ii) We perform the calculation in a static limit at $Q=0$ and $\Omega_m=0$ assuming that the vertex is not singular in this limit. The generalization to the limit of finite $Q,\Omega_m$ is carried out in the next section where spin-fluctuation propagator is calculated.  

To this end, the spin operator vertex can be represented, as all other objects encountered so far, as follows 
\begin{align}
[S_z]_{\alpha\beta} = \sum_{\mu \nu \lambda} M^{S_z}_{\mu \nu \lambda} [ \tau^{\mu} \rho^{\nu} \sigma^{\lambda} ]_{\alpha\beta}.
\end{align}
At the bare level, namely without disorder lines, we have
\begin{align}\label{Sz_0}
[S_z]^{0}_{\alpha\beta} = \sum_{\mu \nu \lambda} M^{0}_{\mu \nu \lambda} [ \hat{\tau}_{\mu} \hat{\rho}_{\nu} \hat{\sigma}_{\lambda} ]_{\alpha\beta}\, , \quad M^{0}_{\mu \nu \lambda} = \delta_{\mu,1} \delta_{\nu,0}\delta_{\lambda,3}.
\end{align}
Then the RPA resummation gives, see Fig. \eqref{Fig:Sz-Ladder},
\begin{align}\label{Ladder_Sz}
M^{S_z}_{\mu \nu \lambda} = M^{0}_{\mu \nu \lambda} + \sum_{\mu'\nu'\lambda'}M^{S_z}_{\mu' \nu' \lambda'} P_{\mu' \nu' \lambda'| \mu'' \nu'' \lambda''}\Gamma_{\mu'' \nu'' \lambda''|\mu \nu \lambda} 
\end{align}
where the matrix elements of $\hat{\Gamma}$ are given by \eqref{dec17a} and polarization operator matrix $\hat{P}$ has its own representation of the type specified by Eq. \eqref{T-dis} as 
\begin{equation}\label{Eq:Pi-munugamma}
[P_{Q,\Omega_m}]_{\mu\nu\gamma|\mu'\nu'\gamma'}=\sum_k\tr\left[(\hat{\tau}_\mu\hat{\rho}_\nu\hat{\sigma}_\gamma)\hat{G}_+(\hat{\tau}_{\mu'}\hat{\rho}_{\nu'}\hat{\sigma}_{\gamma'})\hat{G}_-\right].
\end{equation}
Here Green's function $\hat{G}_{\pm}=\hat{G}_{k\pm Q/2,\omega_n\pm\Omega_m/2}$ should be taken from Eq. \eqref{GpwnDis}. We observe that owing to the index structure specified by Eq. \eqref{Sz_0}, we only need the matrix elements of the polarization operator $P_{\mu' \nu' \lambda'|\mu'' \nu'' \lambda''}$ with at least one of the sets of indices being $103$. In this case, the only such nonvanishing matrix element is the diagonal one as given by an expression at $(Q,\Omega_m)\to0$
\begin{equation}\label{Eq:Pi103}
[P_{\omega_n}]_{103|103}=-\frac{8\pi\nu_F}{\sqrt{\Delta^2_\omega+\varpi^2_n}}
\end{equation}
that follows from the calculation of the corresponding trace at zero temperature and $M=0$. As a result, the $S_z$ vertex renormalization does not produce any other vertices. In other words, the equation \eqref{Ladder_Sz} is not a matrix but a scalar equation that is trivially solved by using $[\hat{\Gamma}]_{103|103}=(u^2_0+u^2_\pi)/8$ from Eq. \eqref{dec17a} and Eq. \eqref{Eq:Pi103}
 \begin{equation}\label{Eq:VS-dressed}
 M^{S_z}_{\mu \nu \lambda} 
 = \delta_{\mu,1} \delta_{\nu,0}\delta_{\lambda,3}V_S, \quad V_S(\omega_n)= 
  \frac{ 1 }{ 1 + \Gamma_{\text{t}}/D_\omega}.
 \end{equation}
In writing the last expression we used notations from Eqs. \eqref{Eq:Gamma-tm} and \eqref{Eq:D}. Returning back to one of the initial assumptions, we notice that indeed the vertex renormalization is infrared regular, and we were correct in assuming that it can be computed right at the transition and at zero $Q$ and $\Omega_m$. 

In short, the calculations carried out in this section are done within the leading ladder approximation of impurity diagram technique, as crossing of impurity lines in diagrams lead to an extra smallness in a parameter $\Gamma_{0,\pi}/E_F\ll1$, and in addition we did not include the mixed (interference) scattering terms $\propto u_0u_\pi$. 


\section{Spin-fluctuation propagator}\label{App:Propagator}

The SDW state breaks SU(2) spin invariance and as a result the spectrum of collective excitations in the ordered and disordered phases is different. In the paramagnetic phase all three spin polarization directions contribute equally to the thermodynamic properties. In the ordered state, two of them are turned into the Goldston modes of the transverse spin fluctuations while the longitudinal mode hardens away from the QCP. Without loss of generality we assume that the spin order is along the $z$ direction. 

To derive an expression for the spin-fluctuation propagator we need a more accurate expression for the polarization operator at finite momenta and frequencies. The expansion we need to deal with in Eq. \eqref{Eq:Pi-munugamma} reads
\begin{equation}\label{Eq:PiQOmega}
\hat{P}_{Q,\Omega_m}\approx\hat{P}_{\omega_n}+\delta\hat{P}^{[Q^2]}_{\omega_n}Q^2+\delta\hat{P}^{[\Omega^1]}_{\omega_n}\Omega_m+\delta\hat{P}^{[\Omega^2]}_{\omega_n}\Omega^2_m,
\end{equation}
where we used the matrix form of presentation. It is obvious that the linear in $Q$ term vanishes upon an angular averaging implicit in the trace of Eq. \eqref{Eq:Pi-munugamma}. As the first task, we compute nonvanishing matrix elements of individual terms and find 
\begin{subequations}
\begin{align}
&\big[\delta\hat{P}^{[Q^2]}_{\omega_n}\big]_{103|103}=\frac{\pi\nu_Fv^2_F}{(\Delta^2_\omega+\varpi^2_n)^{3/2}}, \\
&\big[\delta\hat{P}^{[\Omega^1]}_{\omega_n}\big]_{103|213}=\frac{4\pi\nu_F(\Delta_\omega\varpi'_{n}-\Delta'_\omega\varpi_n)}{(\Delta^2_\omega+\varpi^2_n)^{3/2}},\\
&\big[\delta\hat{P}^{[\Omega^2]}_{\omega_n}\big]_{103|103}=\frac{\pi\nu_F}{(\Delta^2_\omega+\varpi^2_n)^{5/2}}\nonumber\\ 
&\times\big[(\Delta^2_\omega+\varpi^2_n)(\Delta_\omega\Delta''_\omega+\varpi_n\varpi''_n)+3(\Delta_\omega\varpi'_{n}-\Delta'_\omega\varpi_n)^2\big].
\end{align}
\end{subequations}
In addition observe that $\big[\delta\hat{P}^{[\Omega^1]}_{\omega_n}\big]_{213|103}=-\big[\delta\hat{P}^{[\Omega^1]}_{\omega_n}\big]_{103|213}$. As the next step, we focus on the calculation of the full resummed disorder averaged polarization operator defined by \begin{equation}
\hat{\Pi}_{Q,\Omega_m}=T\sum_{\omega_n}\big[\hat{P}_{Q,\Omega_m}(1-\hat{\Gamma}\circ\hat{P}_{Q,\Omega_m})^{-1}\big],
\end{equation}
where $\hat{\Gamma}$ is specified by Eq. \eqref{dec17a}. In the main text, we used a simplified notation, 
\begin{equation}\label{Eq:Pz-vs-P103}
\Pi^z_{Q,\Omega_m}\equiv[\hat{\Pi}_{Q,\Omega_m}]_{103|103},
\end{equation}
see Eq. \eqref{Eq:Pi}, and similarly for $P^z_{Q,\Omega}$. Being interested in the low energy limit specified by an expansion of Eq. \eqref{Eq:PiQOmega}, we need to re-expand $\hat{\Pi}_{Q,\Omega_m}$ in powers of $Q$ and $\Omega_m$ in order to establish the resulting expression for the spin-fluctuation propagator defined by Eq. \eqref{Eq:L},  
\begin{equation}\label{SPL}
L_{Q,\Omega_m}=\frac{1}{\pi\nu_F}
\left(\gamma+\frac{Q^2}{Q_c^2}+\frac{\Omega^2_m}{\Omega_c^2}\right)^{-1}. 
\end{equation}
It is relatively straightforward to determine zeroth order and $Q^2$ order terms in $L_{Q,\Omega_m}$ as they come from the diagonal matrix elements of $\hat{\Pi}_{Q,\Omega_m}$. However, the $\propto \Omega^2_m$ term is more complicated as it receives corrections from the off-diagonal elements as well. For instance, a product of two linear in $\Omega_m$ terms of the form 
\begin{equation}
\big[\delta P^{[\Omega^1]}_{\omega_n}\big]_{103|\mu\nu\lambda}\Gamma_{\mu\nu\lambda|\mu'\nu'\lambda'}\big[\delta P^{[\Omega^1]}_{\omega_n}\big]_{\mu'\nu'\lambda'|103}
\end{equation}
with implicit summation over repeated indices, contributes to $\Omega^2_m$ order for a combination of indices $(\mu,\nu,\lambda)=213$ since $[\hat{\Gamma}]_{213,213}=-(u^2_0-u^2_\pi)/8$ as it follows from Eq. \eqref{dec17a}. It is crucial though that all linear in $\Omega_m$ terms cancel from $L_{Q,\Omega_m}$.    

Below we list the corresponding expressions for the parameters in the paramagnetic and SDW-ordered state. Furthermore, in order to establish a connection to the quasiclassical analysis, and to make all expressions manifestly real, we redefine superconducting anomalous $-if_\omega=f'_{\omega}\to f_\omega$ and magnetic $is_\omega=s'_\omega\to s_\omega$ Green's functions. In these notations 
\begin{align}
\varpi_n=\omega_n+\Gamma_{\text{t}}g_{\omega_n}, \quad \Delta_{\omega}=\Delta+\Gamma_{\text{m}}f_{\omega_n}, \nonumber \\
M_{\omega}=M-\Gamma_{\text{t}}s_{\omega_n},\quad  g^2_\omega+f^2_\omega+s^2_\omega=1.
\end{align}

\subsection{Paramagnetic state $M=0$}
The explicit computation shows that in the paramagnetic state, the parameters $\gamma$, $Q_c$, and $\Omega_c$ are given by
\begin{subequations}\label{ParamsM=0}
\begin{align}
&\gamma=\frac{1}{\pi\nu_Fg_{\textrm{sdw}}}-8T\sum\limits_{\omega_n}\frac{1}{D_\omega+\Gamma_{\text{t}}}, \\
&Q_c^{-2}=v_F^2T\sum\limits_{\omega_n}\frac{1}{D_\omega\left[D_\omega+\Gamma_{\text{t}}\right]^2}, \\
&\Omega_c^{-2}=T\sum\limits_{\omega_n}\frac{1}{D_\omega\left[D_\omega+\Gamma_{\text{t}}\right]^2}\times\\&
\left[\Delta_\omega\Delta_\omega''+\varpi_n\varpi_n''+\frac{(\Delta_\omega\varpi_n'-\varpi_n\Delta_\omega')^2}{D^2_\omega}\left[3+\frac{2\Gamma_{\text{m}}}{D_\omega}\right]\right],
\end{align}
\end{subequations}
where $\varpi_n''$ and $\Delta_\omega''$ denote the derivatives with respect to the Matsubara frequency. 

\subsubsection{Quantum critical point}

The quantum critical point is found by setting $\gamma=0$:
\begin{equation}\label{QCP}
{\frac{1}{g_{\textrm{sdw}}}=
\sum\limits_{\omega_n}\frac{\pi\nu_FT}{\sqrt{(\omega_n+\Gamma_{\text{t}}g_{\omega_n})^2+\left(\Delta+\Gamma_{\text{m}}f_{\omega_n}\right)^2}+\Gamma_{\text{t}}}}.
\end{equation}
Note that the value of the coupling constant $g_{\textrm{sdw}}$ and the values of disorder are fixed in the mean-field theory. 
Alternatively, we can solve this equation to determine the dependence of the dimensionless coupling constant 
$\pi\nu_Fg_{\textrm{sdw}}$ as a function of $\Gamma_{\text{t}}$ with the fixed ratio $\Gamma_\pi/\Gamma_0$ keeping in mind the $\Delta$ must be computed using the mean-field equations. 

Let us check that Eq. (\ref{QCP}) is consistent with the mean-field equations (\ref{BulkMF}). To do that, we first recall the self-consistency equation for the SDW order parameter:
\begin{align}\label{sdwmf}
\frac{M}{g_{\textrm{sdw}}}=\pi\nu_FT\sum\limits_{\omega_n}^\Lambda s_{\omega_n} 
=\pi\nu_FT\sum\limits_{\omega_n}^\Lambda \frac{Mg_{\omega_n}}{\omega_n+2\Gamma_{\text{t}}g_{\omega_n}}.
\end{align}
Comparing this expression with Eq. (\ref{QCP}) it makes sense to express the denominator in Eq. (\ref{sdwmf}) in terms of the denominator in Eq. (\ref{QCP}). With the help of the mean-field equations, we have $(\Delta/f_\omega)+(\Gamma_0-\Gamma_\pi)=(M/2s_\omega)+(\Delta/2f_\omega)-\Gamma_\pi$. 
Next, we express the ratio $M/s_\omega$ using the mean-field equation
$M/s_\omega=(\omega/g_\omega)+2\Gamma_{\text{t}}$,
and invoke the fact that at the QCP $M=0$ so that $g_\omega^2+f_\omega^2=1$. This gives 
\begin{align}\label{aux5}
&\left[\Delta+(\Gamma_0-\Gamma_\pi)f_\omega\right]^2+\left(\omega+\Gamma_{\text{t}}g_\omega\right)^2=\nonumber \\ 
&\left(\omega+\Gamma_{\text{t}}g_\omega\right)^2\left(1+\frac{f_\omega^2}{g_\omega^2}\right)=\frac{\left(\omega+\Gamma_{\text{t}}g_\omega\right)^2}{g_\omega^2}.
\end{align}
With this relation in hand, we go back to the denominator in Eq. (\ref{QCP}):
\begin{align}\label{denomsdwmf}
&\sqrt{(\omega_n+\Gamma_{\text{t}}g_{\omega_n})^2+\left(\Delta+\Gamma_{\text{m}}f_{\omega_n}\right)^2}+\Gamma_{\text{t}}=\nonumber \\ 
&\frac{\omega_n+\Gamma_{\text{t}}g_{\omega_n}}{g_{\omega_n}}+\Gamma_{\text{t}}=\frac{\omega_n+2\Gamma_{\text{t}}g_{\omega_n}}{g_{\omega_n}}.
\end{align}
Inserting this expression into Eq. (\ref{QCP}) gives Eq. (\ref{sdwmf}).

\subsubsection{Derivatives} 

In a similar fashion one may verify that the other two expansion coefficients $Q_c$ and $\Omega_c$ in Eq. \eqref{SPL} obtained from quasiclassical equations coincide with that obtained via resummation of disorder averaged diagrams. For that to check one needs derivatives of Eilenberger function that appear after momentum and energy expansion.  We note that the derivatives of the functions $g_{\omega_n}$ and $f_{\omega_n}$ can be actually expressed in terms of functions themselves. Indeed, 
combining the mean-field equation with the normalization condition, the function $g_{\omega_n}$ is determined by the root of the equation
\begin{equation}\label{eq4gwn}
\frac{1}{g_{\omega_n}^2}=1+\frac{\Delta^2}{(\omega_n+2\Gamma_\pi g_{\omega_n})^2}, 
\end{equation}
so for derivatives we obtain
\begin{subequations}\label{eq4gwn12}
\begin{align}
&g_{\omega_n}'\!=\!
\frac{\Delta^2g_{\omega_n}^3}{(\omega_n+2\Gamma_\pi g_{\omega_n})^3-2\Gamma_\pi\Delta^2g_{\omega_n}^3}=
\frac{f_{\omega_n}^3}{\Delta-2\Gamma_\pi f_{\omega_n}^3}, \\
&g_{\omega_n}''\!=-3\left(\frac{g_{\omega_n}'}{g_{\omega_n}}\right)^2\!\!(g_{\omega_n}-\omega_ng_{\omega_n}')f_{\omega_n}^2=-\frac{3\Delta g_{\omega_n}f_{\omega_n}^8}{(\Delta-2\Gamma_\pi f_{\omega_n}^3)^3}
\end{align}
\end{subequations}
where we repeatedly used the quasiclassical equations. 
We note that the second derivative of $g_{\omega_n}$ is actually negative for the arbitrary values of the Matsubara frequencies and disorder. 
Similarly it follows:
\begin{subequations}\label{fderives}
\begin{align}
f'_{\omega_n}&=-\frac{g_{\omega_n}}{f_{\omega_n}}g_{\omega_n}'=-\frac{g_{\omega_n}f_{\omega_n}^2}{\Delta-2\Gamma_\pi f_{\omega_n}^3}, \\ 
f''_{\omega_n}&=-\frac{(g'_{\omega_n})^2}{f_{\omega_n}}
-\frac{g_{\omega_n}}{f_{\omega_n}}g_{\omega_n}''+\frac{g_{\omega_n}}{f_{\omega_n}^2}f'_{\omega_n}g_{\omega_n}'\nonumber\\
&=-\frac{g_{\omega_n}}{f_{\omega_n}}g_{\omega_n}''-\frac{(g'_{\omega_n})^2}{f_{\omega_n}^3}.
\end{align}
\end{subequations}

\subsection{SDW state $M\not=0$}
The expression for the quantum critical parameter $\gamma$ in the magnetically ordered case near QCP is
\begin{align}\label{gammaSDW}
\gamma(M)&=\gamma(M=0)+12T\sum\limits_{\omega_n}\frac{M_\omega^2}{D_\omega\left[D_\omega+\Gamma_{\text{t}}\right]^2}\nonumber \\ 
&\times\left[1+\frac{2}{3}\left(\frac{\Gamma_{\text{t}}\Delta^2_\omega}{D^3_\omega+\Gamma_{\text{t}}\varpi^2_n}-
\frac{\Gamma_{\text{t}}\varpi^2_n}{D^3_\omega+\Gamma_{\text{t}}\Delta^2_\omega}\right)\right]
\end{align}
It was derived perturbatively in smallness of $M\ll\Delta$. This proves our earlier assertion, made in the previous section, that finite $M$ introduces only regular corrections. The expressions for the $Q_c$ and $\Omega_c$ can also be found in a similar manner.


\section{Quantum-fluctuation corrections}\label{App:Fluctuations}

\subsection{Heat capacity}

The contribution of the quantum critical spin fluctuations to the free energy is given by Eq. \eqref{Eq:FQF}, which in the explicit notations reads 
\begin{equation}\label{Eq1}
\frac{\delta F_{\textrm{QF}}}{T}=\frac {\mathcal{N}}{2}\sum\limits^{+\infty}_{m=-\infty}\int\frac{d^2Q}{(2\pi)^2}\ln\left[L^{-1}_{Q,\Omega_m}\right].
\end{equation}
To perform here the Matsubara summation we first single out the $m=0$ term and reduce the remaining summation over the positive Matsubara frequencies. Thus we find 
\begin{align}\label{AEq1}
&\frac{\delta F_{\textrm{QF}}}{T}=\frac{\mathcal{N}}{2}\int\frac{d^2Q}{(2\pi)^2}\ln(\pi\nu_F E_Q^2/\Omega^2_c)\nonumber \\ 
&+\mathcal{N}\int\frac{d^2Q}{(2\pi)^2}\ln\left[
\prod\limits_{n=1}^\infty\left(1+\frac{E_Q^2}{\Omega_n^2}\right)\prod\limits_{m=1}^\infty
\left(\frac{\pi\nu_F\Omega_m^2}{\Omega_c^2}\right)\right],
\end{align}
where we introduced a notation $E_Q=\Omega_c\sqrt{\gamma+(Q/Q_c)^2}$. The first term in the last equation represents a zero-point-motion correction and, therefore, does not produce the temperature dependent contribution to the heat capacity. In the second term, there are two products under the logarithm which we discuss separately. The first product can be evaluated simply by using Mittag-Leffler's theorem from the theory of meromorphic functions in the complex analysis  
\begin{equation}\label{AEq2}
\prod\limits_{m=1}^\infty\left(1+\frac{E_Q^2}{\Omega_m^2}\right)=\frac{\sinh(\pi z)}{\pi z}, 
\quad z=\frac{E_Q}{2\pi T}.
\end{equation}
To evaluate the second product in Eq. \eqref{AEq1}, which is formally divergent, we need to use a regularization scheme to assign it a finite value. This can be done by employing the zeta-function regularization procedure well known in the context of path-integral representation of the statistical mechanics. It is based on the following functional determinant formula 
\begin{equation}
\Det O=\exp\left[-\left.\frac{d\zeta_O(s)}{ds}\right|_{s=0}\right],\quad \zeta_O(s)=\sum_m\frac{1}{\lambda^s_m}
\end{equation}
where $\lambda_m$ refers specifically to the eigenvalues of an operator $O$. For our purposes, we can simply notice that $\Omega^2_m$ in the second product under the logarithm in Eq. \eqref{AEq1}, can be understood as eigenvalues of the differential operator $O=-\partial^2_\tau$ in imaginary time. As a result, with the values of the Riemann zeta function $\zeta(0)=-1/2$ and $\zeta'(0)=-(1/2)\ln(2\pi)$ this recipe yields the following result:
\begin{equation}\label{AEq3}
\sum\limits_{m=1}^\infty
\ln\left(\frac{\pi\nu_F\Omega_m^2}{\Omega_c^2}\right)\doteq\ln\left(\frac{\Omega_c}{\sqrt{\pi\nu_F}T}\right),
\end{equation}
where symbol $\doteq$ emphasizes equality in the sense of the above specified regularization. 
Thus, dropping all the temperature independent contributions, the expression for the fluctuation correction to the free energy is
\begin{equation}\label{AEq4}
\frac{\delta F_{\textrm{QF}}}{T}=\mathcal{N}\int\frac{d^2Q}{(2\pi)^2}\ln\left[1-e^{-E_Q/T}\right].
\end{equation}
At temperatures $T>\Delta_{\text{QCP}}$ with $\Delta_{\text{QCP}}=\Omega_c\sqrt{\gamma}$ we can simply take $E_Q$ at $\gamma\to0$, and using then a tabulated integral 
\begin{equation}
\int^{\infty}_{0}x\ln[1-e^{-x}]dx=-\zeta(3)    
\end{equation}
obtain a fluctuation correction to the heat capacity in the form of Eq. \eqref{C-QF} from the main text. In the opposite limit of extremely low temperatures, $T<\Delta_{\text{QCP}}$, an asymptotic analysis of momentum integral in Eq. \eqref{AEq4} reveals an exponential suppression of heat capacity $\delta C_{\text{QF}}\propto e^{-\Delta_{\text{QCP}}/T}$.   

\begin{figure}
    \centering
    \includegraphics[width=3.5in]{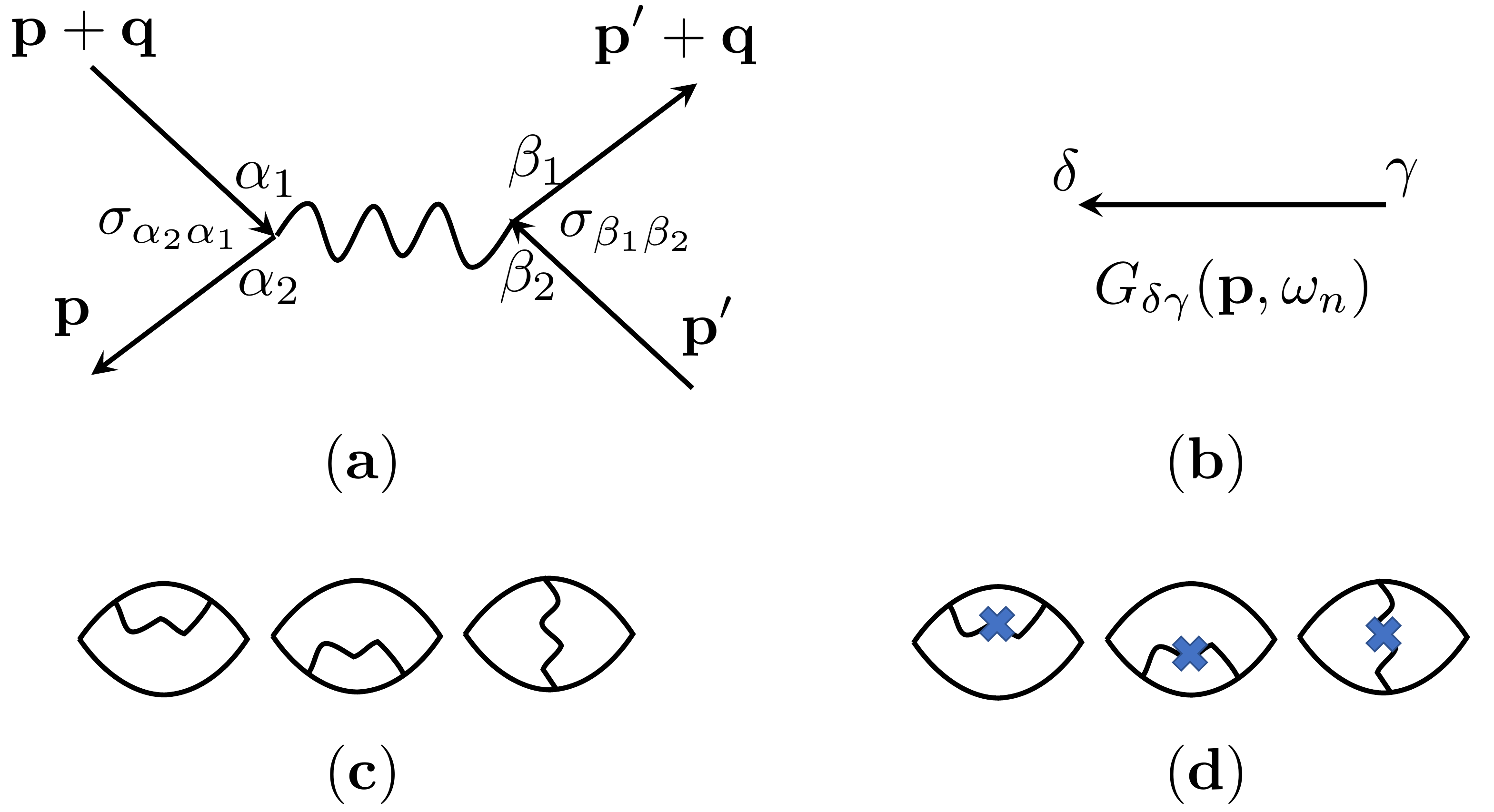}
    \caption{Panel (a) shows the convention for putting indices on the interaction line in the spin sector. Panel (b) shows the convention for indices on the single particle Green's function. Panel (c) depicts three diagrams that contribute to the electromagnetic response kernel $\delta K_{\text{QF}}$. The first two are the density of states ($F_{\text{DOS}}$) contributions, while the last is the Maki-Thompson ($F_{\text{MT}}$) term. Panel (d) replicates leading diagrams to extract the most divergent part where one has to take the derivative $\partial_\gamma$ of the
propagators marked by a symbol of a bold cross (\textbf{x}) placed on the (wavy) interaction lines.}
    \label{Fig:One-loop}
\end{figure}

\subsection{London penetration depth}

The fluctuation correction to the inverse square of the London penetration depth is given by the static current-current correlation function
\begin{equation}\label{invlam2}
\delta\left(\lambda^{-2}\right)=\lim\limits_{q\to 0}\delta\left\langle \mathbf{ j}_\perp(q,\omega=0)\mathbf{j}_\perp(-q,\omega=0)\right\rangle,
\end{equation}
where the transversal component $\mathbf{j}_\perp$ of the current operator is defined by
\begin{equation}\label{jtrans}
\mathbf{j}_\perp(q)=\frac{e}{m}\sum\limits_{\bm{p}\sigma}\bm{p}_\perp\left(
\hat{c}^\dagger_{\bm{p}+\frac{\bm{q}}{2}\sigma}\hat{c}_{\bm{p}-\frac{\bm{q}}{2}\sigma}-
\hat{f}^\dagger_{\bm{p}+\frac{\bm{q}}{2}\sigma}{\hat{f}}_{\bm{p}-\frac{\bm{q}}{2}\sigma}\right),
\end{equation}
where $\bm{p}_\perp=\bm{p}-\hat{\bm{q}}\left(\bm{p}\cdot\hat{\bm{q}}\right)$. At the one-loop level the correction is graphically depicted by a sum of three diagrams, see Fig. \ref{Fig:One-loop}, representing the self-energy and vertex corrections, respectively, that contain four Green's functions and one fluctuation propagator. From the current-current correlation function, on the paramagnetic side of the transition where $M=0$ and $\mathcal{N}=3$, we find

\begin{equation}\label{deltaK}
\left(\frac{\partial \delta K_{\textrm{QF}}}{\partial\gamma}\right)_{M=0}=-\frac{\mathcal{N}}{2}e^2v^2_F
T\sum\limits_{\Omega_m}\int\frac{d^2Q}{(2\pi)^2}\frac{\partial L_{Q,\Omega_m}}{\partial \gamma}F_{\text{l}},
\end{equation}
where 
\begin{subequations}
\begin{equation}
F_{\text{l}}=F_{\text{DOS}}+F_{\text{MT}},
\end{equation}
\begin{align}
&F_{\text{DOS}}=2T\sum_{\omega_n}V^2_S(\omega_n)\nonumber \\ &\times\int\frac{d^2\bm{p}}{(2\pi)^2}\tr\left[\hat{G}_{\bm{p},\omega_n}\hat{\tau}_3\hat{G}_{\bm{p},\omega_n}\hat{\tau}_3\hat{G}_{\bm{p},\omega_n}\hat{\Xi}^z\hat{G}_{\bm{p},\omega_n}\hat{\Xi}^z\right],   
\end{align}
\begin{align}
&F_{\text{MT}}=T\sum_{\omega_n}V^2_S(\omega_n)\nonumber \\ &\times\int\frac{d^2\bm{p}}{(2\pi)^2}\tr\left[\hat{G}_{\bm{p},\omega_n}\hat{\tau}_3
\hat{G}_{\bm{p},\omega_n}\hat{\Xi}^z\hat{G}_{\bm{p},\omega_n}\hat{\tau}_3\hat{G}_{\bm{p},\omega_n}\hat{\Xi}^z\right],    
\end{align}
\begin{equation}
V_S(\omega_n)=\left[1+\frac{\Gamma_0+\Gamma_\pi}{\sqrt{\Delta^2_\omega+\varpi^2_n}}\right]^{-1},
\end{equation}
\end{subequations}
with $\hat{\Xi}^z=\hat{\tau}_1\hat{\rho}_0\hat{\sigma}_3$ and $\tr[\ldots]$ representing matrix trace only. The expression for the fermionic loop $F_{\text{l}}$ was simplified by neglecting Green's function dependence on the bosonic frequencies $\Omega$ and momenta $Q$. This is justified in the low-temperature limit, as fermions are fully gapped while bosons are soft. For this reason we were able to disentangle integrations over the fermionic and bosonic modes in $K_{\text{QF}}$. We have further differentiated the response kernel over $\gamma$ to ensure it is well convergent in the ultraviolet, unlike the bare bubble. This enables us to freely set an incoming momentum to zero, $q=0$, from the very beginning.

To proceed further with the analysis of these expressions we take the matrix trace and convert $d^2\bm{p}$ integral into $d\xi_{\bm{p}}$ with the density of states $\nu_F$. This way one arrives at $F_{\text{l}}$ for $T\to0$ in the form 
\begin{align}
&F_{\text{l}}=6\nu_F\int^{+\infty}_{-\infty}\frac{V^2_S(\omega)\Delta^2_\omega d\omega}{(\Delta^2_\omega+\varpi^2_\omega)^{5/2}}\nonumber \\ &=6\nu_F
\int^{+\infty}_{-\infty}\frac{\Delta^2_\omega }{\big(\Delta^2_\omega+\varpi^2_\omega\big)^{3/2}}
\frac{d\omega}{\big(\sqrt{\Delta^2_\omega+\varpi^2_\omega}+\Gamma_0+\Gamma_\pi\big)^2}.
\end{align}
In can be shown that at $T\ll\Delta$ finite-$T$ corrections to $F_\text{l}$ are exponentially small, $e^{-\Delta/T}\ll1$, thus negligible.  
With this result at hand, the fluctuation correction $\delta K_{\text{QF}}$ can be rewritten in the form
\begin{equation}
\frac{\partial}{\partial\gamma}\left[\frac{\delta K_{\text{QF}}}{K_0}\right]=-\frac{\mathcal{N}\Omega^4_c}{\pi\nu^2_F}F_{\text{l}}\int\frac{d^2Q}{(2\pi)^2}T\sum^{+\infty}_{m=-\infty}\frac{1}{[E^2_Q+\Omega^2_m]^2}. 
\end{equation}
The bosonic frequency Matsubara sum evaluates to a simple expression 
\begin{align}
&T\sum^{+\infty}_{m=-\infty}\frac{1}{[E^2_Q+\Omega^2_m]^2}\nonumber \\ &=\frac{1}{4E^3_Q}
\left[\coth(E_Q/2T)+\frac{E_Q}{2T}\frac{1}{\sinh^2(E_Q/2T)}\right].
\end{align}
In order to extract the leading $T$ asymptote in the regime above the QCP gap, namely when $T>\Delta_{\text{QCP}}$, it is sufficient to expand the last expression at small argument $E_Q/T\ll1$. As a result 
\begin{equation}
\frac{\partial}{\partial\gamma}\left[\frac{\delta K_{\text{QF}}}{K_0}\right]=-\frac{\mathcal{N}T}{4\pi^2\nu^2_F}F_{\text{l}}
\int^{\infty}_{0}\frac{QdQ}{[\gamma+(Q/Q_c)^2]^2}. 
\end{equation}
The remaining integrals are elementary and give 
\begin{equation}
\frac{\delta K_{\text{QF}}}{K_0}=-\frac{\mathcal{N}TQ^2_c}{8\pi^2\nu^2_F}F_{\text{l}}\ln\left(\frac{1}{\gamma}\right).
\end{equation}
To reproduce the result quoted in the main text, we evaluate $Q_c$ and $F_{\text{l}}$ at QCP where $\Gamma_{0,\pi}\ll\Delta$, thus to the main order $v_FQ_c=\sqrt{\pi}\Delta$ and $F_{\text{l}}=8\nu_F/\Delta^2$, which gives 
\begin{equation}
\frac{\delta K_{\text{QF}}}{K_0}=-\frac{\mathcal{N}}{4}\frac{T}{E_F}\ln\left(\frac{1}{\gamma}\right).
\end{equation}
In the opposite limit, $T<\Delta_{\text{QCP}}$, we have $\coth(E_Q/2T)\to1$, so that $\partial_\gamma\delta K_{\text{QF}}\propto\int QdQ/E^{3}_Q\propto1/\sqrt{\gamma}$, and finally $\delta K_{\text{QF}}/K_0\simeq -\sqrt{\gamma}(\Delta/E_F)$.   

\begin{figure}
    \centering
   \includegraphics[width=0.45\textwidth]{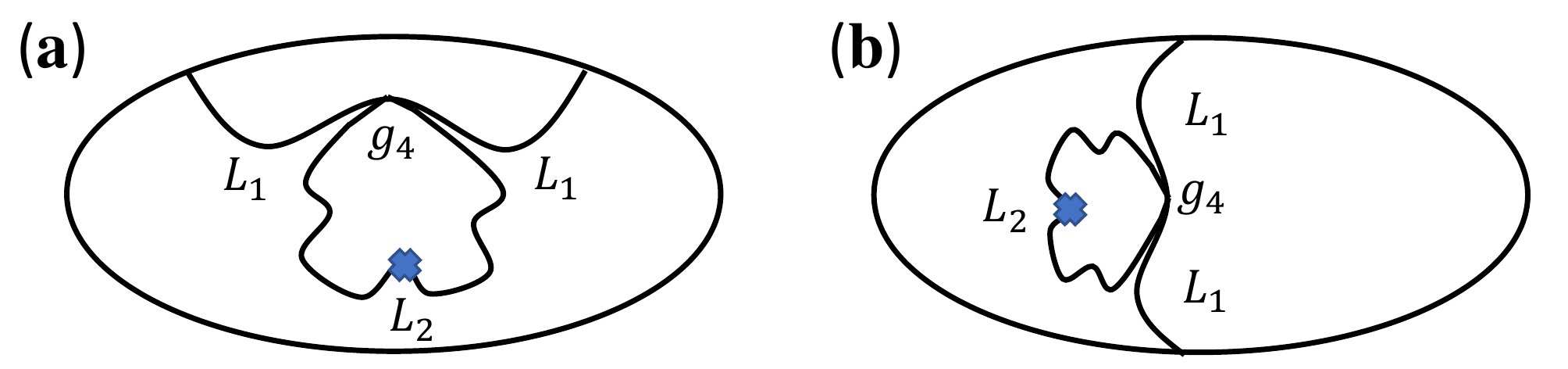}
    \caption{The exemplary contributions to the electromagnetic response at the two-loop level.
    Diagram (a) denotes the density of states contribution (and the mirror image diagram is implicit), 
    while diagram (b) denotes the Maki-Thompson correction. 
    The bold cross (\textbf{x}) in the diagram indicates the derivative with respect to $\gamma$. 
    Applying the derivative to the other two propagators amounts to differentiating the propagator with
   the mass renormalization included. }
  \label{Fig:Two-loop}
\end{figure}

\subsection{Two-loop estimates}

In this section we focus on the corrections to the penetration depth due to the nonlinear terms of spin fluctuations. In the boson action they are captured by the following interaction term  
\begin{align}\label{2L11}
&A_4[S^z(\bm{q},\omega)]= g_4 \sum_{\bm{q}_1,\bm{q}_2,\bm{q}_3} T^3\nonumber \\ 
 &\sum_{\omega_1,\omega_2,\omega_3}  
S^z_{\bm{q}_1,\omega_1} S^z_{\bm{q}_2,\omega_2} S^z_{\bm{q}_3,\omega_3} 
S^z_{-\bm{q}_1-\bm{q}_2-\bm{q}_3,-\omega_1-\omega_2,-\omega_3} 
\end{align}
The goal is to demonstrate the validity of an asymptotic expansion of fluctuation corrections and determine the parameter that controls it.  
To this end, we consider the simplest contribution beyond the one-loop level originating from the boson coupling in Eq. \eqref{2L11}.
Graphically, such contributions are described by the set of three diagrams, Fig.~\ref{Fig:Two-loop}, which correspond to the three diagrams at the level of the first loop shown in Fig.~\ref{Fig:One-loop}(c). 

The fluctuation correction proportional to $g_4$ is more singular than the one-loop corrections considered in the main text. This two-loop correction is given by the expression
\begin{align}\label{2L12}
\delta^{(2)} K_{\textrm{QF}} \propto g_4 (ev_F)^2 F_{\text{l}} 
\int \prod_{i=1}^2\frac{d\Omega_id^2Q_i}{(2\pi)^3} L^2_{Q_1,\Omega_1} L_{Q_2,\Omega_2}\, .
\end{align}
Here again as in the one-loop contribution the integration over $Q_2$ and $\Omega_2$ is ultraviolet divergent, and we apply the same technical step of taking the derivative with respect to $\gamma$. Certainly, the derivative may equally be applied to the propagator $ L^2_{Q_1,\Omega_1}$. This, however, is the same contribution as considered in the main text with the boson propagator including the effect of the mass renormalization. Here we do not consider the boson mass renormalization. For this reason we focus on contribution originating from taking the derivative of the boson propagator not attached to the fermion propagators as shown 
in Fig.~\ref{Fig:Two-loop}. For these reasons we focus on the least singular contribution originating from \eqref{2L11} which reads
\begin{align}\label{2L13}
\partial_{\gamma} \delta^{(2)} K_{\textrm{QF}} \propto g_4 (ev_F)^2 F_{\text{l}}
\int \prod_{i=1}^2\frac{d\Omega_i d^2Q_i}{(2\pi)^3} L^2_{Q_1,\Omega_1} L^2_{Q_2,\Omega_2}
\end{align}
This results in the product of two convergent integrals each very similar to the one we encountered in the one-loop calculation.
For further estimates we take $\Delta$ for $\Omega_c$ and $\Delta/v_F$ for $Q_c$.
Furthermore, the local boson coupling is estimated as $g_4 \propto \nu_F/\Delta^2$.
With this in mind we obtain the next to the leading order term in asymptotic series 
\begin{align}\label{2L14}
\partial_{\gamma} \delta^{(2)} K_{\textrm{QF}}/K_0 \propto  (\Delta/E_F)^2 \gamma^{-1}.
\end{align}
This estimate leads to the conclusion that we have the following hierarchy of contributions,
\begin{align}\label{2L15}
\partial_{\gamma} \delta^{(2)} K_{\textrm{QF}}/\partial_{\gamma} \delta^{(1)} K_{\textrm{QF}} \sim \partial_{\gamma} \delta^{(1)} K_{\textrm{QF}}/K_0  \propto \left(\frac{\Delta}{E_F}\right) \gamma^{-1/2}
\end{align}
where $\delta^{(1)} K_{\textrm{QF}}$ is the one-loop contribution considered in the main text.
This confirms that the critical region is defined by the condition $\gamma \approx (\Delta/E_F)^2$.

\end{document}